\documentclass[twocolumn,showpacs,amsmath,amssymb,pre]{revtex4}

\usepackage{graphicx}
\usepackage{dcolumn}
\usepackage{bm}

\begin{document}

\title{Relaxation in a glassy binary mixture: Mode-coupling-like power laws, dynamic heterogeneity
and a new non-Gaussian parameter}

\author{Elijah Flenner}
\author{Grzegorz Szamel}
\affiliation{Department of Chemistry, Colorado State University, Fort Collins, CO 80525}

\begin{abstract}
We examine the relaxation of the Kob-Andersen
Lennard-Jones binary mixture using Brownian dynamics computer simulations.  
We find that in accordance with mode-coupling theory the self-diffusion coefficient
and the relaxation time show power-law dependence on temperature. However,
different mode-coupling temperatures and power laws can be obtained from the
simulation data depending on the range of temperatures chosen for the power-law
fits. The temperature that is commonly reported as this system's mode-coupling transition temperature, 
in addition to being obtained from a power law fit, 
is a crossover temperature at which there is a change in the
dynamics from the high temperature homogeneous, diffusive relaxation
to a heterogeneous, hopping-like motion. The hopping-like motion is evident
in the probability distributions of the logarithm of single-particle displacements:
approaching the commonly reported mode-coupling temperature these distributions start exhibiting two peaks. 
Notably, the temperature at which the hopping-like motion appears for the smaller particles
is slightly higher than that at which the hopping-like motion appears for the larger ones.
We define and calculate a new non-Gaussian parameter
whose maximum occurs
approximately at the time at which the two peaks in 
the probability distribution of the logarithm of displacements are most evident.

\end{abstract}

\date{\today}


\maketitle

\section{\label{intro}Introduction}

Computer simulations have added a great deal to our knowledge of
relaxation in supercooled liquids. A seminal series of 
studies was performed by Kob and Andersen \cite{KobAndersen} who were interested
in comparing the results of molecular dynamics simulations 
to the predictions of the mode-coupling theory \cite{Gotze}.
They observed that as the temperature was lowered the mode-coupling theory
gave a good qualitative description of the relaxation of the liquid.  
Specifically, they found that the long-time self-diffusion coefficient
and a characteristic relaxation time showed power law behavior as 
$T_c = 0.435$ was approached. Since their investigation, 
this temperature has been referred to as the mode-coupling transition temperature
for the Kob-Andersen Lennard-Jones mixture.      
Later work \cite{Gleim,Kobsr} demonstrated that there
is no true vanishing of the self-diffusion coefficient or true divergence of the 
$\alpha$ relaxation time at $T_c$.  
This is similar to what is observed experimentally
for most glass formers \cite{Shankar}.  
There exists power law like behavior of the relaxation time 
as a temperature $T_c$ is approached and close to $T_c$ there
is a crossover to a different relaxation scenario which extends 
to the laboratory glass transition $T_g$, defined as the temperature at which the
viscosity equals $10^{13}$ poise \cite{Lunkenheimer,Shankar,Ken}. 

Many simulations performed since the Kob and Andersen study have 
resulted in the following relaxation scenario in
supercooled liquids around the crossover temperature $T_c$ 
(see Ref.~\cite{GlotzerNC} and references therein). 
The motion of particles in a slightly supercooled liquid is homogeneous and 
the self part of the van Hove correlation function \cite{HansenMcDonald} is approximately Gaussian
at all times. At lower temperatures the self part of the van Hove
correlation function deviates significantly from Gaussian, the motion 
of the particles is strongly heterogeneous, and, on an intermediate time scale
(up to so-called exchange time), the particles can
be separated by their individual relaxation times.    
It is generally believed that at short times the particles are 
confined to cages of neighboring particles and a ``cooperative motion'' \cite{Donati1} 
of particles has to occur to facilitate cage escape \cite{KobDH, Donati2}.  
The timescale of this cooperative motion increases
with decreasing temperature. In the original studies \cite{KobDH,Donati1} it was assumed that this
time scale is around the peak in the non-Gaussian parameter
$\alpha_2(t) = \frac{3}{5} \left< \delta r^4 \right>/\left<\delta r^2 \right>^2  - 1$,
where $\delta r$ is the distance over which a particle moved in time $t$.
It has been accepted that the peak position of $\alpha_2(t)$
``roughly locates the time of maximum dynamic heterogeneity'' \cite{Reichman}. 
While this is a natural first choice, there is no a priori reason
to choose this time. It has been shown that the van Hove correlation
function deviates strongly from a Gaussian distribution over a much longer time scale \cite{SzamelFlenner}.  
Moreover, it is also possible to identify particles which 
remain slower than other particles for time much longer than 
the time scale given by the peak position of $\alpha_2(t)$ \cite{FlennerSzamel}.
We should point out that, although the time scale of the peak position of $\alpha_2(t)$ is 
often singled out in connection with dynamic heterogeneity, it has also been recognized 
that there exists dynamic heterogeneity on a much longer time scale, and a four-point correlation
function that is sensitive to this longer-time heterogeneity 
has been defined and investigated (see Ref. \cite{Lacevic}
and references therein). 

In this work we examine the relaxation of a supercooled liquid
using Brownian dynamics computer simulations.  Experiments conducted 
on Brownian systems (\textit{i.e.} on colloidal suspensions)
have so far failed to show any significant departure from the power law behavior of the
self-diffusion coefficient and the characteristic relaxation time 
\cite{Megen}. 
Thus, it was suggested that the predictions of the 
original (also known as ``idealized'') mode-coupling theory provide a good description of the
relaxation of the fluid.  However, it has since been demonstrated using computer simulations 
that the relaxation of a supercooled fluid is the same for 
Newtonian dynamics \cite{KobAndersen}, stochastic dynamics \cite{Gleim},
and Brownian dynamics \cite{SzamelFlenner}.  
In particular, in computer simulations departures from mode-coupling-like
power laws and emergence of hopping-like motion have been observed for all three microscopic dynamics. 
The qualitative difference between experimental results
(which do not find deviations from mode-coupling-like power laws) and theoretical ones (which do)
remains unexplained.

The focus of this work is the crossover from the high temperature 
homogeneous relaxation of the slightly supercooled
fluid to the low temperature relaxation around $T_c$. We show that, while the mode-coupling
transition temperature cannot be unambiguously determined from the computer simulation results,
the commonly reported temperature is the crossover temperature for two different modes of relaxation.
Furthermore, we define and calculate a time-dependent function that vanishes identically 
for a Gaussian diffusion process. This new function
$\gamma(t) = \frac{1}{3} \left< \delta r^2 \right> \left< 1/\delta r^2 \right> - 1$, 
hereafter called a new non-Gaussian parameter,
has a peak at a time at which the relaxation seems the most heterogeneous. This time 
is always longer than a characteristic decay time of the incoherent intermediate scattering
function \cite{Hansen} that is known as the $\alpha$ relaxation time. Moreover, the
peak position of the new non-Gaussian parameter has temperature dependence very 
similar to that of the $\alpha$ relaxation time.

The paper is organized as follows.  In section \ref{simdetails} we 
briefly describe the simulation.  In section \ref{differentpowerlaws} we present results
for the mean  square displacement, the self-diffusion coefficient, the 
incoherent intermediate scattering functions, and the $\alpha$ relaxation time (a preliminary
report of some of these results appeared in Ref. \cite{SzamelFlenner}).
We show that two different power-law fits can be obtained depending on the range of
temperatures used. In section \ref{distributions} we present results
for the probability distributions of the logarithm of single-particle displacements. 
These distributions are sensitive to hopping-like dynamics. In section \ref{newnong}
we define and present the new non-Gaussian parameter. In section \ref{conclusion} we discuss
the conclusions that can be drawn from
this work.  

\section{\label{simdetails}Simulation Details}

We simulated a binary mixture of 800 particles of type A and
200 particles of type B that was first considered by Kob and Andersen \cite{KobAndersen}. 
Briefly, the interaction potential is 
$V_{\alpha \beta}(r) = 4\epsilon_{\alpha \beta}[
({\sigma_{\alpha \beta}}/{r})^{12} - ({\sigma_{\alpha \beta}}/{r})^6]$,
where $\alpha, \beta \in \{A,B\}$, $\epsilon_{AA} = 1.0$, 
$\sigma_{AA} = 1.0$, $\epsilon_{AB} = 1.5$, $\sigma_{AB} = 0.8$, 
$\epsilon_{BB} = 0.5$, and $\sigma_{BB} = 0.88$.  The
interaction parameters are chosen to prevent crystallization
\cite{KobAndersen}.  The simulations
are performed with the interaction potential cut at 
$2.5\ \sigma_{\alpha \beta}$, and the box length of the cubic simulation cell
is $9.4\ \sigma_{AA}$. Periodic boundary conditions were used.  

We performed Brownian dynamics simulations.  The equation  
of motion for the position of the $i_{th}$ particle of type $\alpha$, $\vec{r}\,_{i}^{\alpha}$, is
\begin{equation}\label{Lang}
\dot{\vec{r}}\,_{i}^{\alpha} = \frac{1}{\xi_0} \vec{F}_i^{\alpha}
+ \vec{\eta}_i(t) ,
\end{equation}
where the friction coefficient of an isolated particle $\xi_0 = 1.0$ and $\vec{F}_i^{\alpha}$
is the force acting on the $i_{th}$ particle of type $\alpha$,
\begin{equation}
\vec{F}_i^{\alpha}= - \nabla_i^{\alpha} \sum_{j \ne i} \sum_{\beta=1}^2
V_{\alpha \beta}\left(\left|\vec{r}\,_i^{\alpha} - \vec{r}\,_j^{\beta} \right| \right)
\end{equation}
with $\nabla_i^{\alpha}$ being the gradient operator with respect to $\vec{r}\,_i^{\alpha}$.
In Eq. (\ref{Lang}) the dot denotes a time derivative, and
the random noise $\vec{\eta}_i$ satisfies the fluctuation-dissipation 
theorem, 
\begin{equation}\label{fd}
\left\langle \vec{\eta}_i(t) \vec{\eta}_j(t') \right\rangle = 
2 D_0 \delta(t-t') \delta_{ij} \mathbf{1}.
\end{equation}
In Eq. (\ref{fd}), the diffusion coefficient $D_0 = k_B T/\xi_0$ where
$k_B$ is Boltzmann's constant and $\mathbf{1}$ is
the unit tensor.  Since the equation of motion allows for diffusive motion
of the center of mass, all the results will be presented relative the
center of mass (\textit{i.e.} momentary positions of all the particles are always 
relative to the momentary position of the center of mass). 
We will present the results in terms of the reduced units
with $\sigma_{AA}$, $\epsilon_{AA}$, $\epsilon_{AA}/k_B$, 
and $\sigma_{AA}^2 \xi_0/\epsilon_{AA}$ being the units of length, energy, 
temperature, and time, respectively.  The mass of both particles are the 
same and equal to 1.0.

The equations of motion, Eq.~\ref{Lang}, were solved using a Heun algorithm with
a small time step of $5 \times 10^{-5}$.  We simulated the temperatures
$T = 0.44$, 0.45, 0.47, 0.5, 0.55, 0.6, 0.8, 0.9, 1.0, 2.0, 3.0, and 5.0. 
We ran a long equilibration run (at least half as long as the production run)
and four production runs at each temperature, except at $T=0.44$
where we ran six production runs. 
The results are an average over the production runs, which were as long
as $6 \times 10^8$ steps long for the lowest temperature studied.  

\section{\label{differentpowerlaws}Mode-coupling-like power laws}
   
The mode-coupling theory predicts a power law vanishing of the 
self-diffusion coefficient and a power law divergence of the characteristic relaxation time at
the mode-coupling transition temperature $T_c$.  In many 
simulations and experiments there is a range of temperatures 
where power laws fit the diffusion coefficient and the
relaxation time well, and the transition
temperature is obtained from fits of these properties to 
functions of the form $a(T-T_c)^{\gamma}$. 
The $T_c$ obtained in this manner is generally 
referred to as the mode-coupling temperature.  
In this section, we will present results for the mean square displacement, the
diffusion coefficient, the self intermediate scattering functions and
the $\alpha$ relaxation time.  Moreover, we will show that 
for the system studied in this work reasonable power 
law fits can be obtained for the diffusion coefficient and 
the $\alpha$ relaxation time for a transition temperature 
different from the usually accepted mode-coupling transition
temperature of $T_c = 0.435$ if a different range of
temperatures is used for the power law fits.

\begin{figure}
	\includegraphics[scale=0.25]{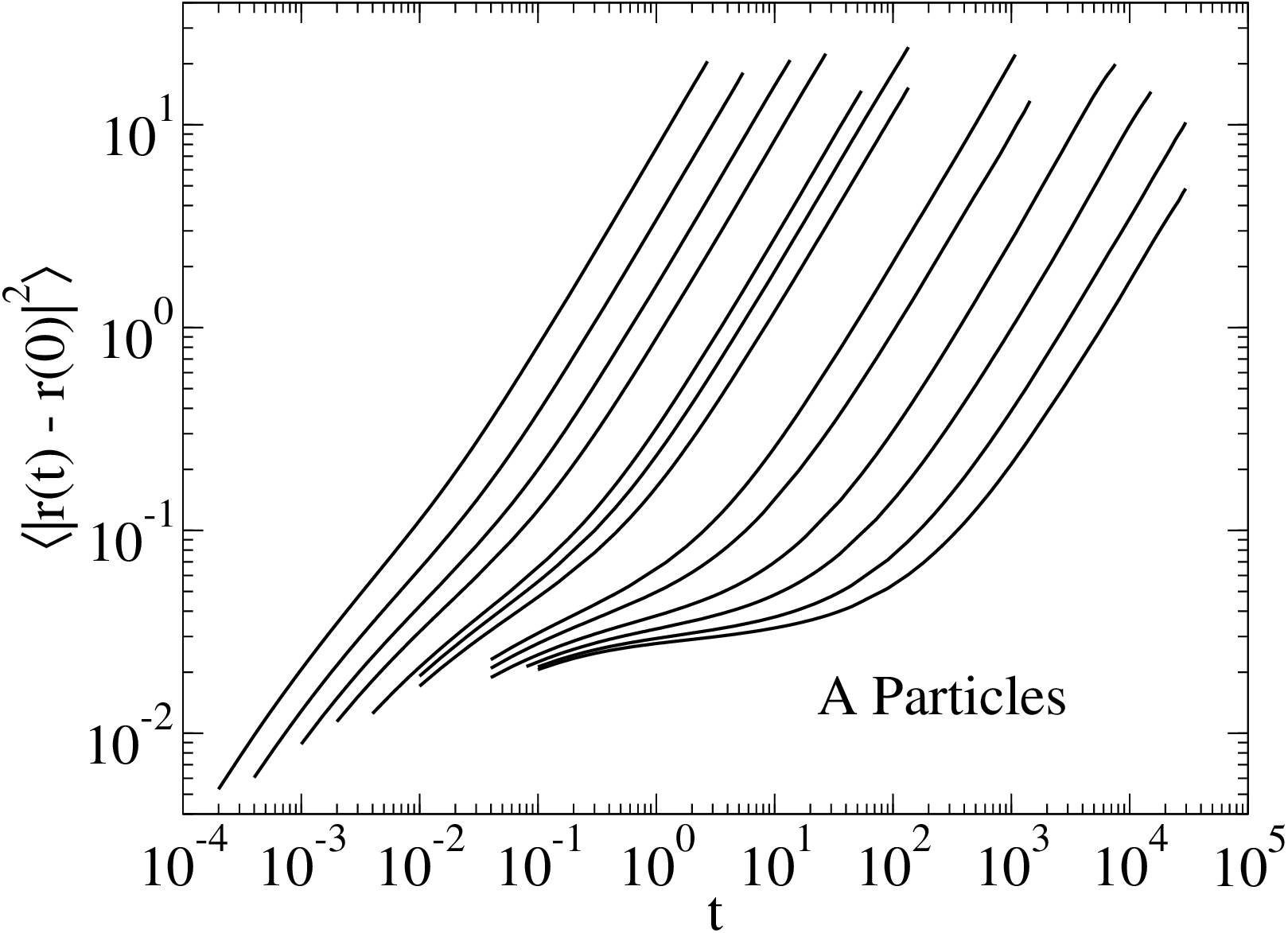}
\caption{\label{msd}The mean square displacement for the A particles
for $T = 5.0$, 3.0, 2.0, 1.5, 1.0, 0.9, 0.8, 0.6, 0.55, 0.50, 0.47, 0.45, 0.44 listed from
left to right.  The graph of the mean square displacement for the B particles is similar.}
\end{figure}
\begin{figure}
	\includegraphics[scale=0.25]{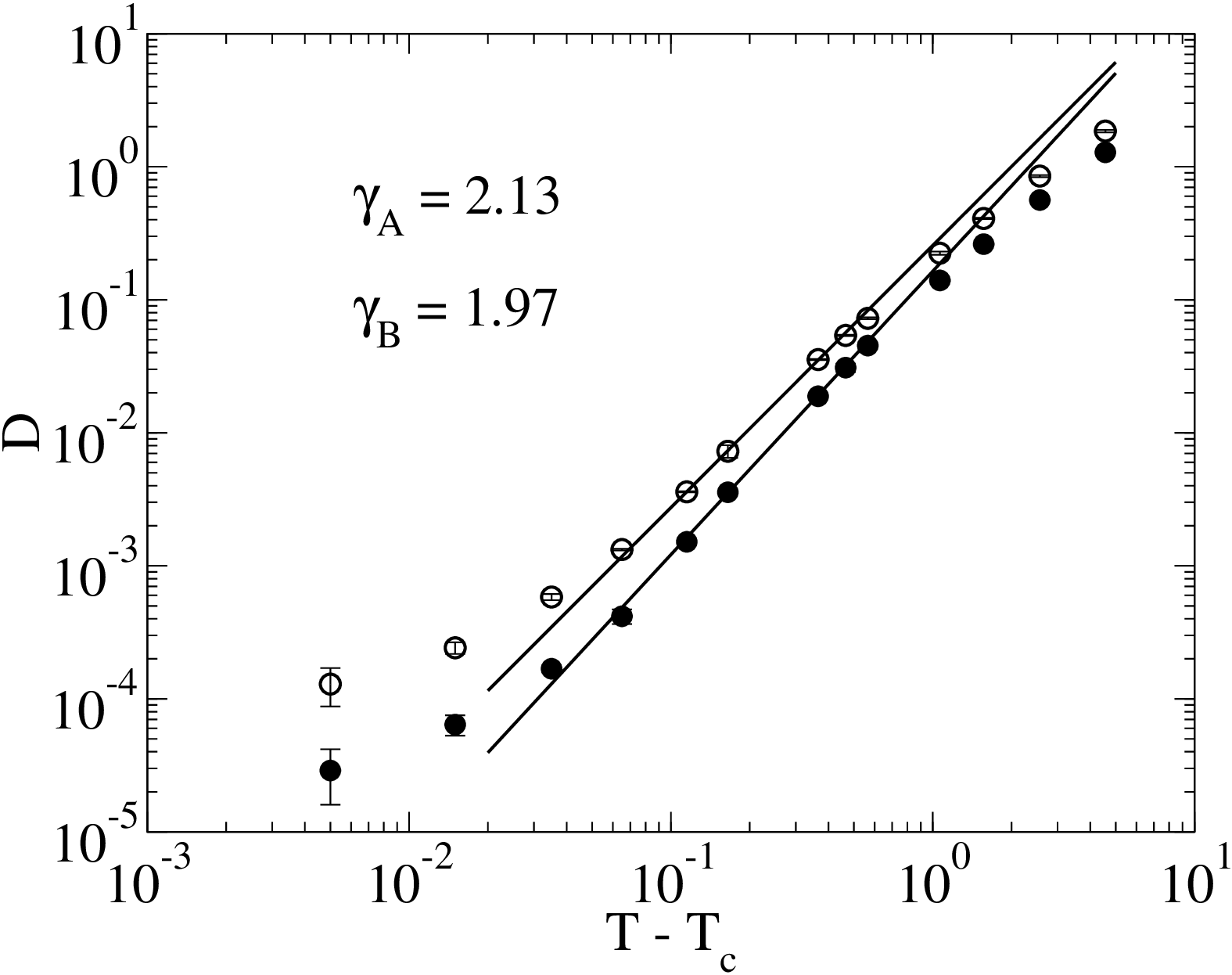}
\caption{\label{Diff}Long time diffusion coefficient for the A (closed symbols)
and the B (open symbols) particles.  The lines are the power law fits 
to the function $a(T-T_c)^{\gamma}$ with $T_c = 0.435$ fixed.}
\end{figure}

Shown in Fig.~\ref{msd} is the single-particle mean square displacement 
$\left< \delta r^2(t) \right> = \left\langle \left| \vec{r}^A(t) - \vec{r}^A(0) \right|^2 \right\rangle$ 
for the A particles.  The graph for the 
B particles looks similar.  
The short time motion is diffusive with a 
temperature dependent diffusion coefficient 
$D_0 = k_B T/\xi_0$ where $k_B$ is Boltzmann's constant
and $\xi_0$ is the friction coefficient (recall that in our units $k_B=1$ and $\xi_0=1$). 
For low temperatures a plateau develops in the log-log
plot at intermediate
times where the mean square displacement does not change
appreciably.  For the lowest temperature the plateau region
spans several decades in time. 
 
The plateau region is generally associated with the
``cage'' effect. It represents a localization of the particles on 
intermediate time scales and has been observed in many 
simulations of glassy systems \cite{KobAndersen,Kobsr}.
Note that there is no
true plateau in the mean square displacement versus time and the 
slope of the mean square displacement versus time 
decreases monotonically.
Therefore there is no inflection point of the mean square
displacement as a function of time.  However, there is an inflection
point in the logarithm of the mean square displacement versus the
logarithm of time.
We use this inflection point to find the cage diameter.
For $T = 0.44$ the inflection point occurs at a value of the
mean square displacement of 0.0288 $\sigma_{AA}$ for the A particles
and 0.0461 $\sigma_{AA}$ for the B particles.
These values of the cage diameter corresponds to a distance
around 0.17 $\sigma_{AA}$ for the A particles 
and around 0.21 $\sigma_{AA}$ for the B particles, which
is much smaller than the diameter of any particle. 
After the plateau, the motion is again diffusive 
with a diffusion coefficient $D < D_0$.    

We determined the long-time
self-diffusion coefficient $D$ from the slope of 
$\left< \delta r^2(t) \right>$ at long times.  The results
for the $A$ and $B$ particles are shown in figure \ref{Diff}.
We observe power law behavior similar to what was 
reported in previous simulations of this system using
Newtonian \cite{KobAndersen} and stochastic dynamics \cite{Gleim,Kobsr}.  
Namely, there is power law 
behavior of the diffusion coefficient for temperatures between 
$T=0.8$ and $T = 0.50$.
Then there are deviations from the power law at and below $T = 0.47$.  
We fit the diffusion coefficients for $0.5 \le T \le 0.8$ 
for the $A$ and $B$ particles to a 
power law of the form $a(T-0.435)^{\gamma}$.  The 
exponents in the fit are given in the figure, and agree 
reasonably well with the exponents found using Newtonian \cite{KobAndersen}
and stochastic dynamics \cite{Gleim,Kobsr,GleimKob}. 

\begin{figure}
	\includegraphics[scale=0.25]{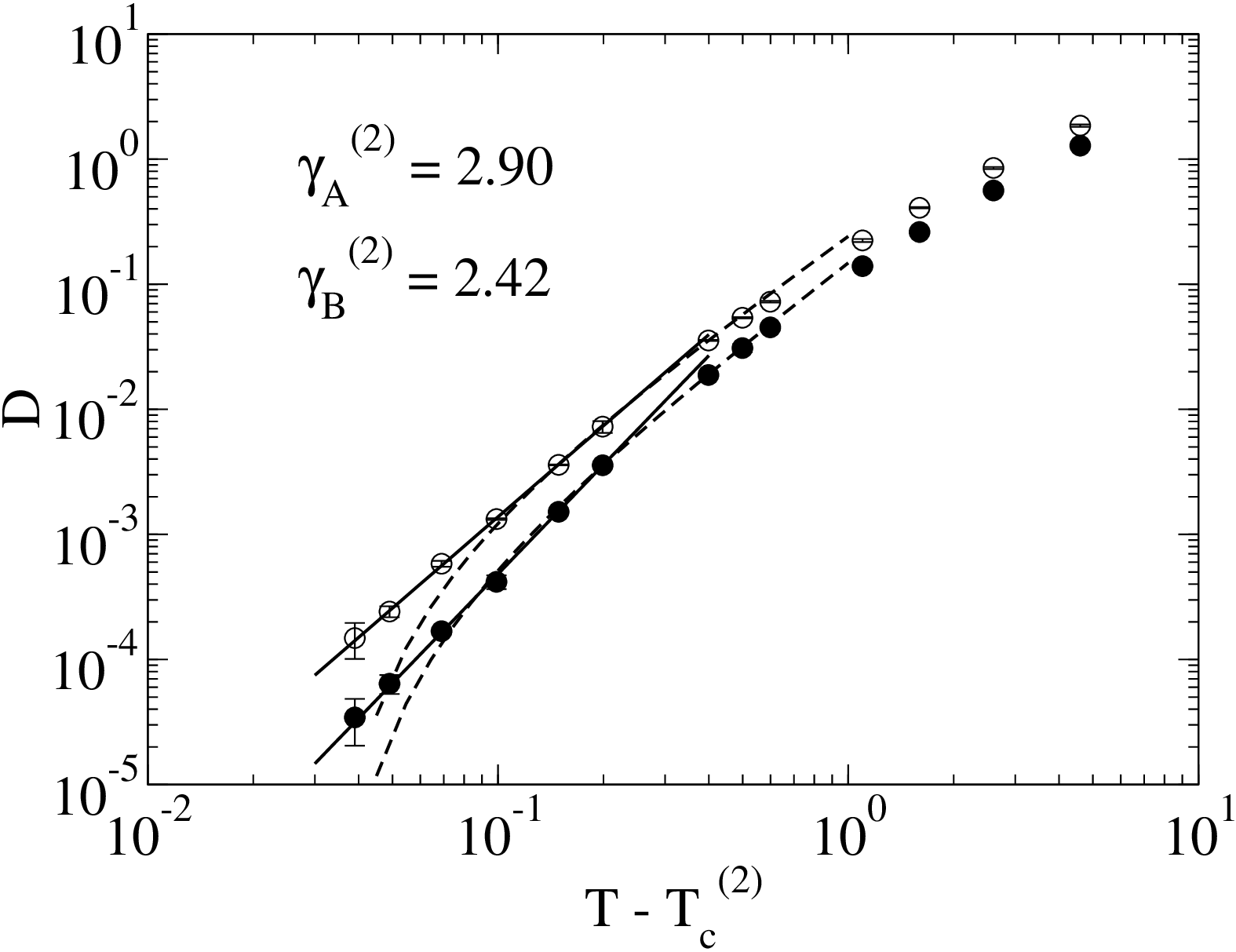}
\caption{\label{Diff1}Diffusion coefficient for the 
A (closed symbols) and B (open symbols) particles.  The solid lines are the ``alternative'' 
power-law fits to the function $a^{(2)}(T - T_c^{(2)})^{\gamma^{(2)}}$ where $T_c^{(2)} = 0.401$.
Dashed lines are fits to the function $a(T - T_c)^{\gamma}$ where $T_c = 0.435$.}
\end{figure}

We also performed three parameter fits to the diffusion coefficient and the 
$\alpha$ relaxation time for different temperature ranges \cite{remark}.  We found that
we can obtain good fits to $D$ and $\tau_{\alpha}$ 
for temperatures $0.44 \le T \le 0.6$.  The transition temperature
depends slightly on the quantity being fitted, and ranges from 
$T_c = 0.391$ to $T_c = 0.409$.  We found the 
average transition temperature obtained from the fits $T_c^{(2)} = 0.401 \pm 0.009$, 
and then fit each quantity to the function
$a^{(2)}(T-0.401)^{\gamma^{(2)}}$. We show this final fit for the self-diffusion coefficient
in Fig.~\ref{Diff1}. Qualitatively, it is clear that the new fit is as good
as the standard fit shown in  Fig.~\ref{Diff}. Quantitatively, we evaluate the quality of
fit by examining $\chi^2 = \sum_i^N \left[ (D(T_i) - y(T_i))/\sigma_i \right]^2$ where
$N$ is the number of data points used in the fit,
$y(T) = a(T-T_c)^{\gamma}$ and $\sigma_i$ 
is the standard deviation of the diffusion coefficient at $T_i$.
For each fit we find the probability $p$ that the value of $\chi^2$ 
should exceed the calculated value by chance given that
the model is correct \cite{mathstat}.  Higher values of $p$ correspond
to a better fit, and the maximum value of $p$ is one.
For the transition temperature $T_c^{(2)} = 0.401$ and the temperature
range $0.44 \le T \le 0.6$, $p = 0.860$ for the A particles
and $p = 0.902$ for the B particles.  This is significantly better than the 
values $p = 0.044$ and $p = 0.0169$ for the A and B particles, respectively, when 
the diffusion coefficient was fit over the temperature range $0.5 \le T \le 0.8$ 
and with $T_c = 0.435$.

\begin{figure}
	\includegraphics[scale=0.25]{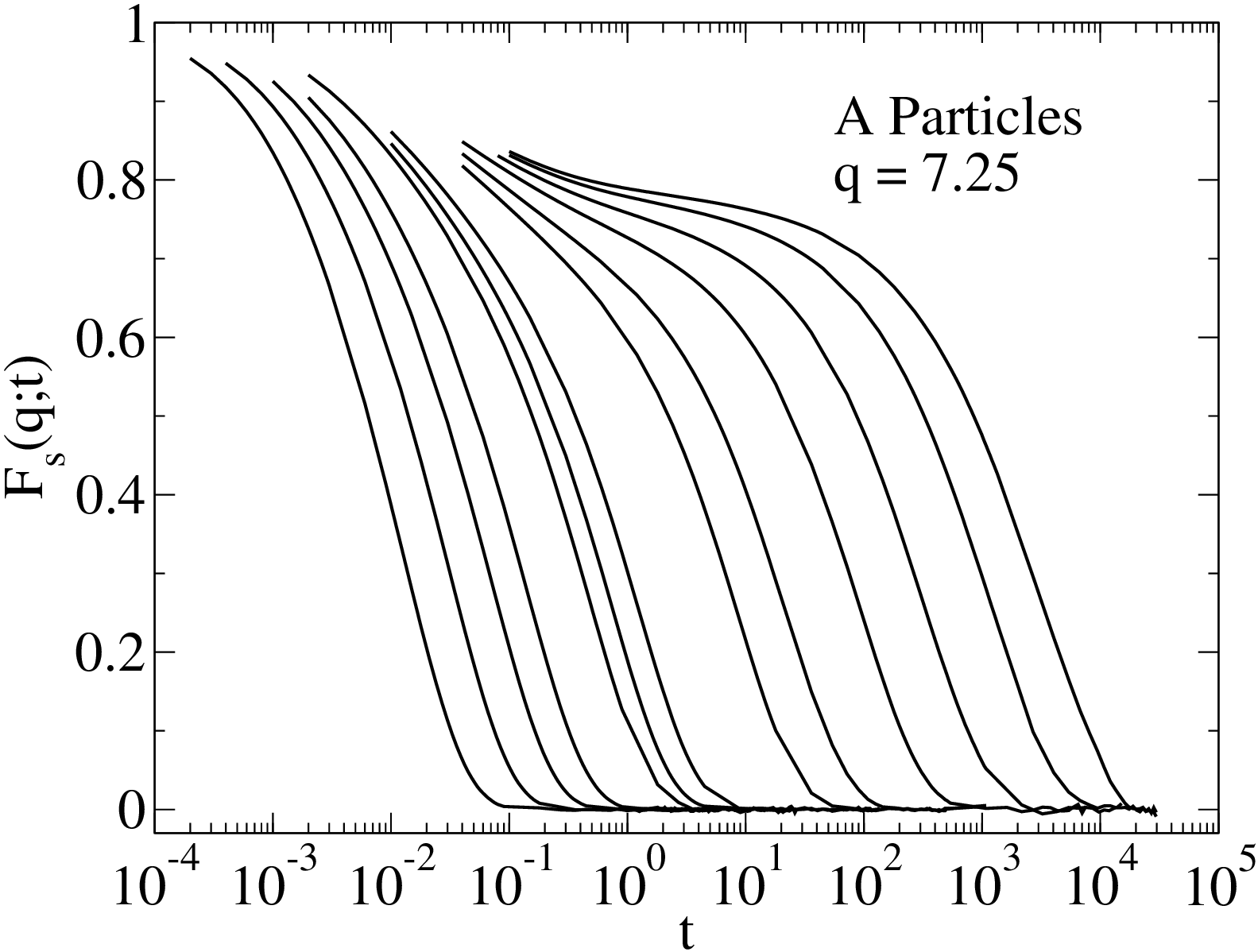}
\caption{\label{scatt}Self intermediate scattering functions for the A particles
for $T = 5.0$, 3.0, 2.0, 1.5, 1.0, 0.90, 0.80, 0.60, 0.55, 0.50, 0.47, 0.45, and 0.44
listed from left to right.  The corresponding graph for the B particles is similar.}
\end{figure}
\begin{figure}
	\includegraphics[scale=0.25]{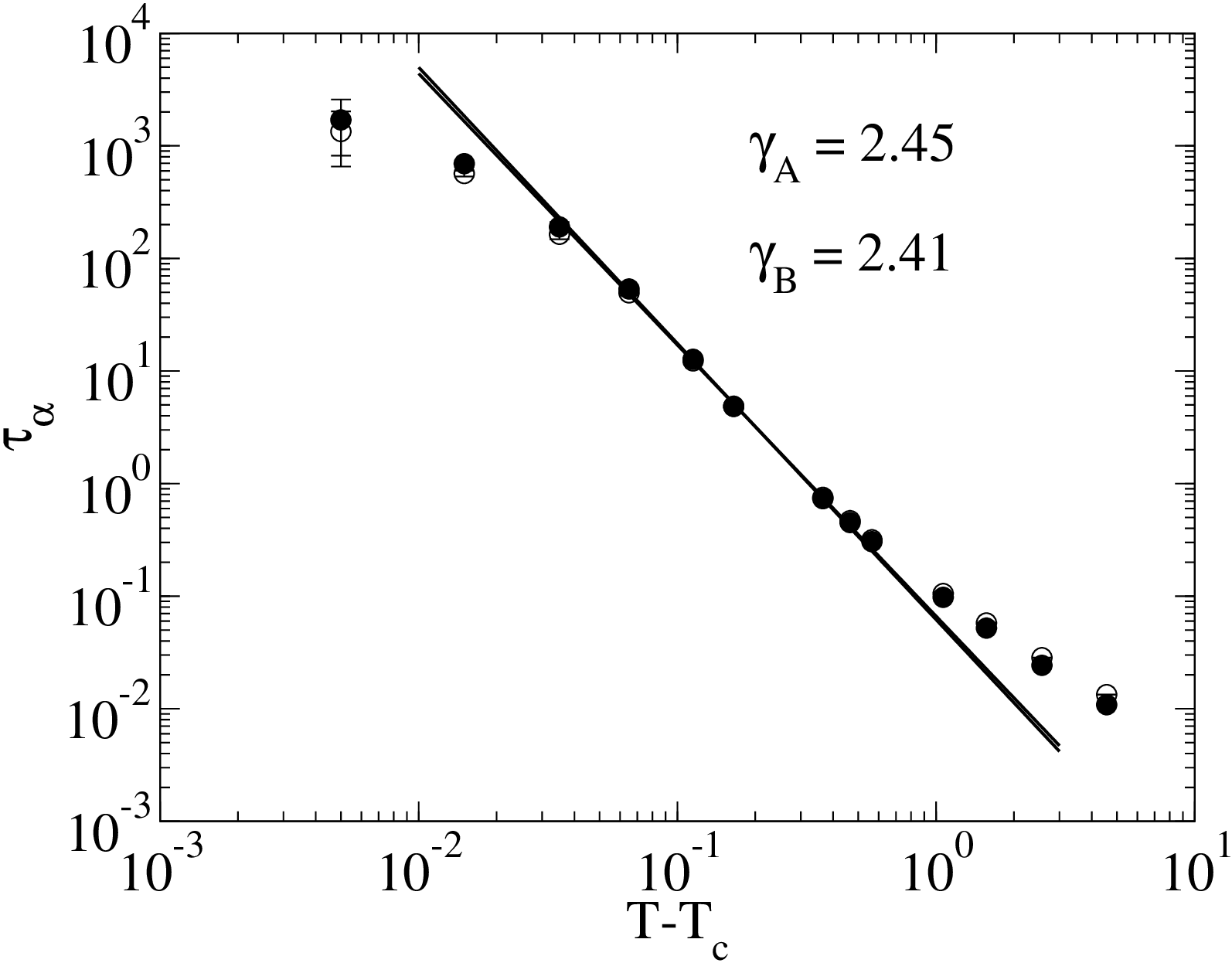}
\caption{\label{tau}The $\alpha$ relaxation time for the A (closed symbols) and B (open symbols)
particles.  The lines are fits to the function $a(T-T_c)^{-\gamma}$ with $T_c = 0.435$ fixed.}
\end{figure}
\begin{figure}
	\includegraphics[scale=0.25]{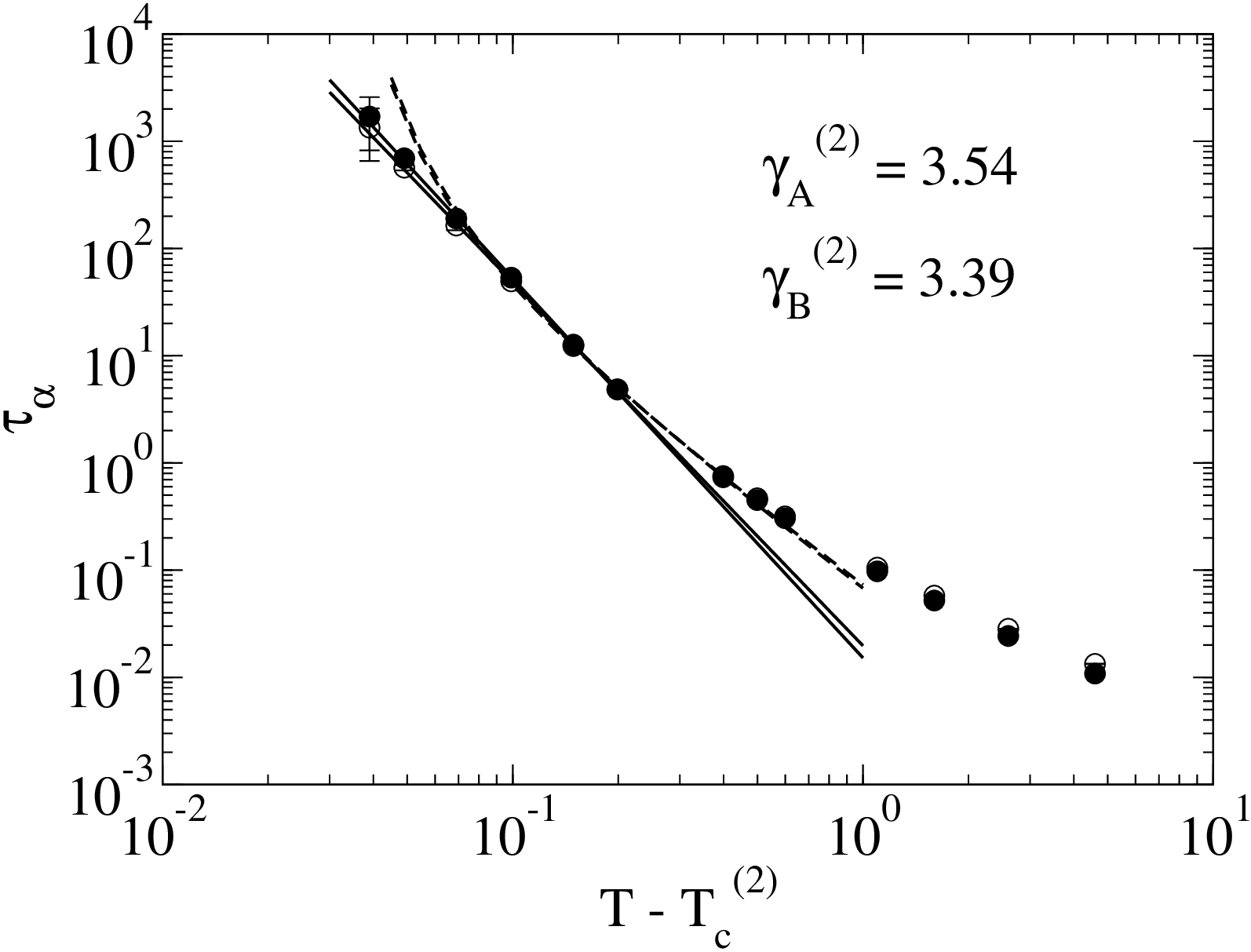}
\caption{\label{tau1}The $\alpha$ relaxation time for the A (closed symbols) and B (open symbols) particles.
The solid line are the ``alternative'' power-law fits
to the function $a^{(2)}(T-T_c^{(2)})^{-\gamma^{(2)}}$ where $T_c^{(2)} = 0.401$. Dashed lines
are fits to the functions $a(T-T_c)^{-\gamma}$ where $T_c = 0.435$.}
\end{figure}

We calculated the self intermediate scattering function 
$F_s^{\alpha}(q,t) = \frac{1}{N_{\alpha}} \left\langle \sum_{j=1}^{N_{\alpha}} 
\exp[i \vec{q} \cdot (\vec{r}\,_j^{\alpha}(t) - \vec{r}\,_j^{\alpha}(0))] 
\right\rangle$, $\alpha\in \{A,B\}$, shown in
Fig.~\ref{scatt}, 
for a wave vector $q = |\vec{q}\,|$ around the first peak of the
partial structure factors for the A $(q=7.25)$ and the
B $(q=5.75)$ particles.  The self intermediate scattering function
decays from its $t=0$ value of one to zero.  
For the same temperatures in which there is a plateau in the 
log-log plot of the mean
square displacement, there is also a plateau in the 
log-log plot of the incoherent intermediate
scattering functions.  The characteristic
time for the decay of the self intermediate scattering 
function is the $\alpha$ relaxation time $\tau_{\alpha}$, which we
define as the time when this scattering function is equal to $1/e$ of its initial value.
It has been observed that other definitions of the $\alpha$ relaxation time
results in the same temperature dependence.  

The $\alpha$ relaxation time is shown in Fig.~\ref{tau}.  
Again there is
power law behavior of the $\alpha$ relaxation time between $0.5 \le T \le 0.8$, 
then there are deviations from the power law for temperatures 
at and below $T = 0.47$.  The lines in the figure are power law fits
to the function $a(T-0.435)^{-\gamma}$ for the temperature range
$0.5 \le T \le 0.8$, which is the same function and the 
same temperature range as for Fig.~\ref{Diff}.
The exponents are given in Fig.~\ref{tau} and are close to what was
found in simulations of the same system using Newtonian \cite{KobAndersen} 
and stochastic dynamics \cite{Gleim,Kobsr} instead of Brownian dynamics.

As explained above, we also fit the $\alpha$ relaxation time to the function
$a^{(2)}(T-0.401)^{-\gamma^{(2)}}$. The result is shown in Fig.~\ref{tau1}.
Again, qualitatively, the new fit is as good
as the standard fit shown in  Fig.~\ref{tau}. Quantitatively, we evaluated
the quality of fits using the same procedure as for the diffusion coefficient.
When $T_c$ was set to 0.401 and we fit the $\alpha$ relaxation
time for $0.44 \le T \le 0.6$, $p=0.598$ for the A particles and
$p = 0.743$ for the B particles.  When the transition temperature
was set to $T_c = 0.435$ and we used the
temperature range $0.5 \le T \le 0.8$,
$p = 0.15$ for the A particles and $p = 0.35$ for the B particles.  
Thus the fits were better for $T_c^{(2)} = 0.401$.

\begin{figure*}
	\includegraphics[scale=0.25]{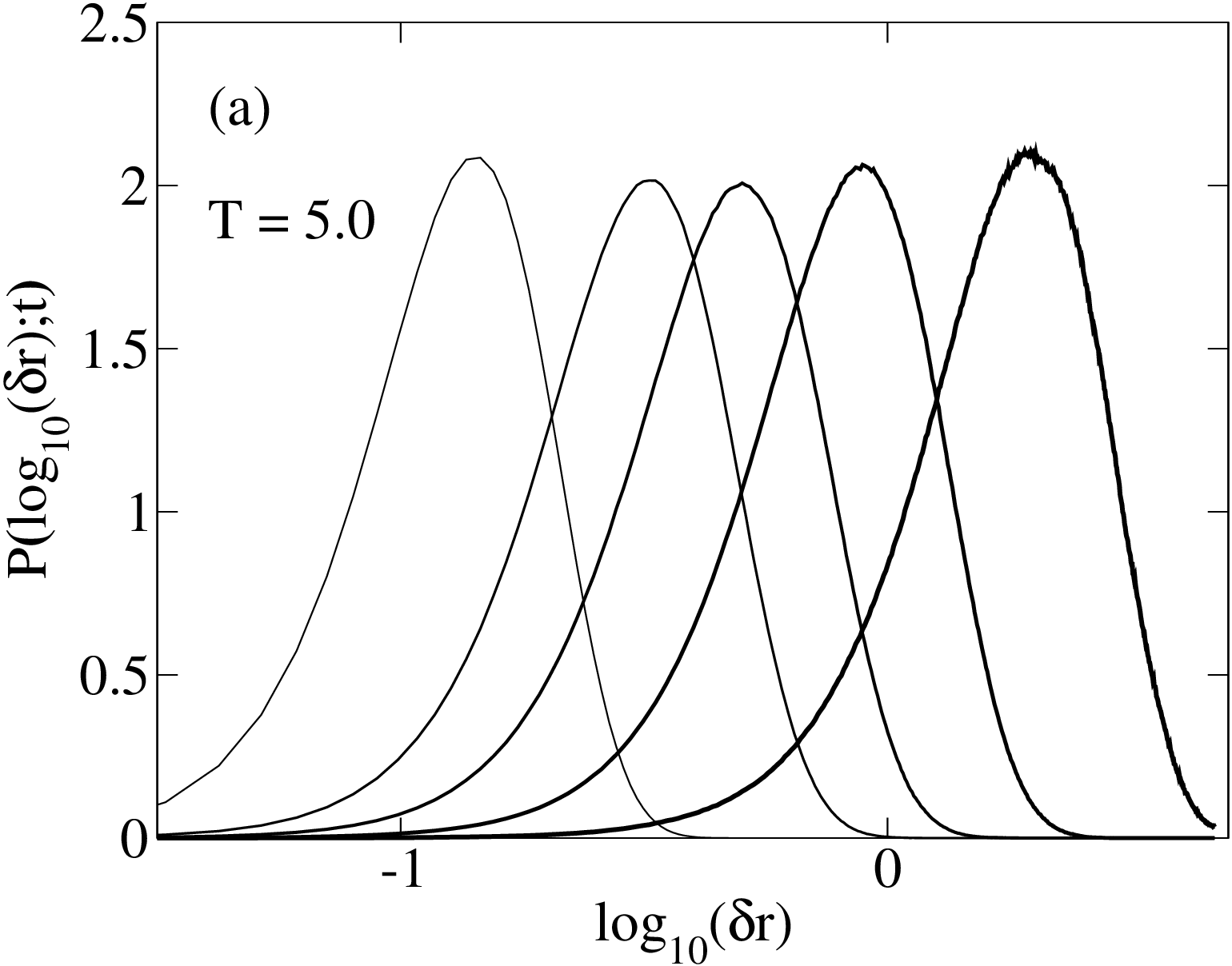}\hspace{1cm}
	\includegraphics[scale=0.25]{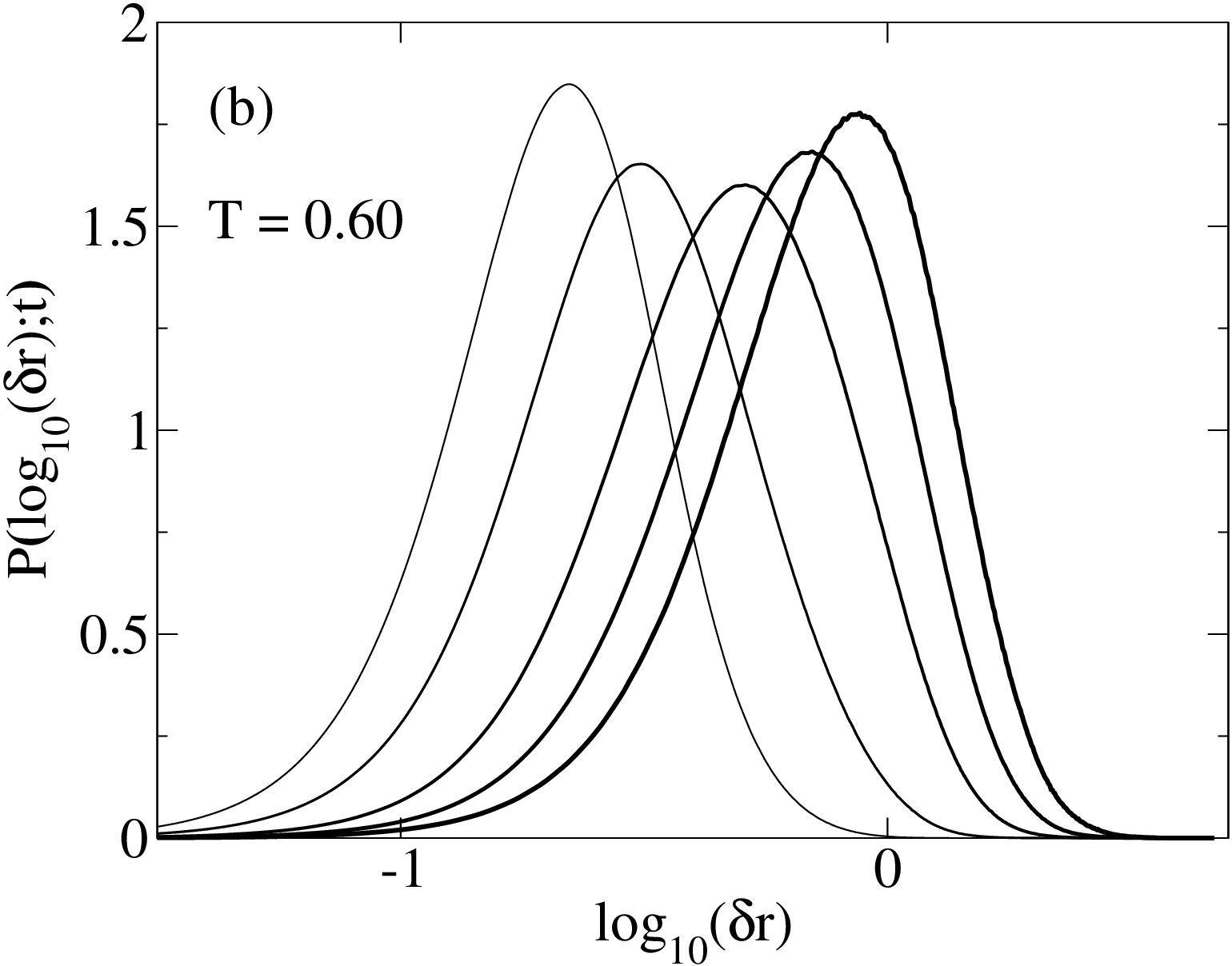}\\[0.25cm]
	\includegraphics[scale=0.25]{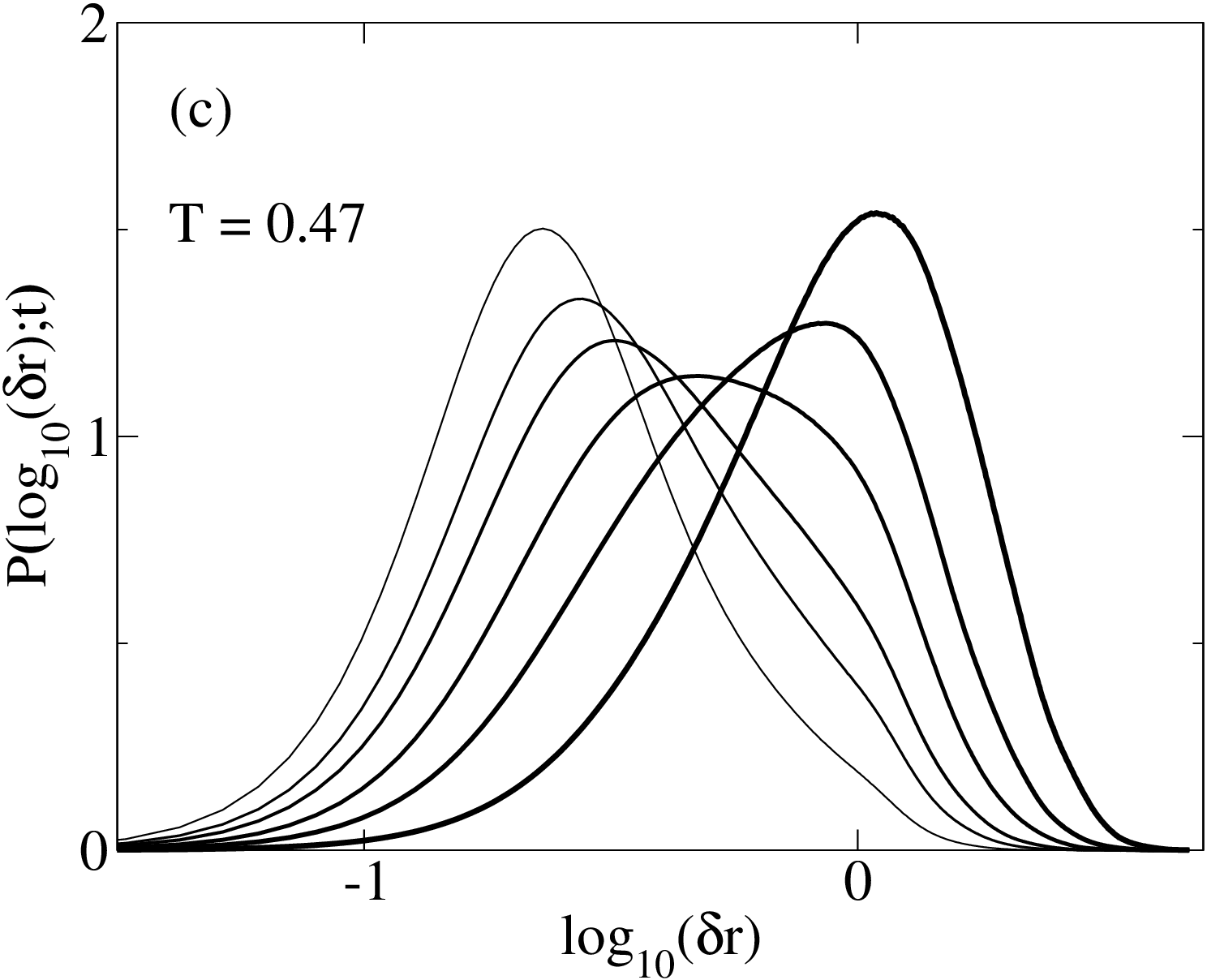}\hspace{1cm}
	\includegraphics[scale=0.25]{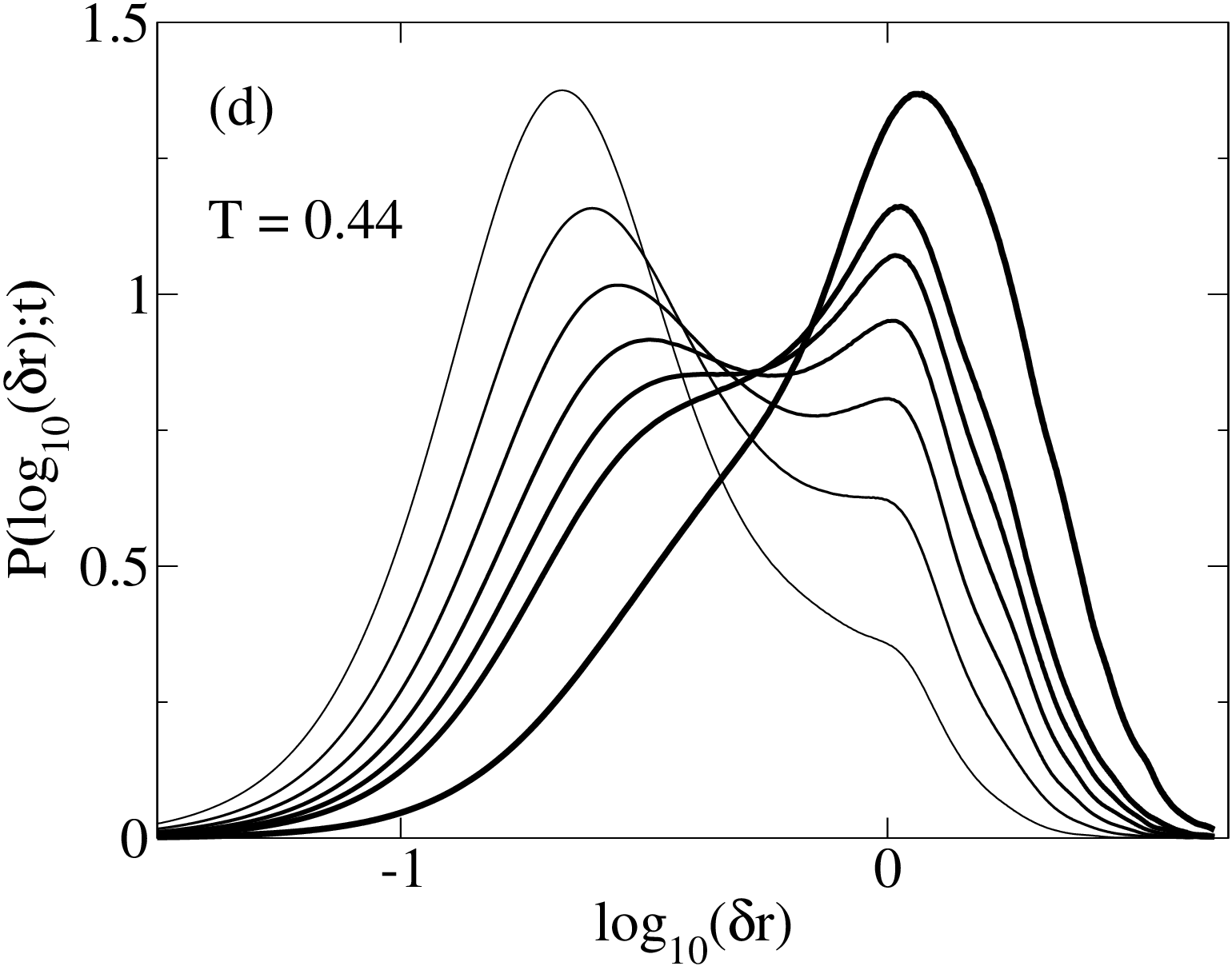}
\caption{\label{vha}The probability of the logarithm of single particle
displacements $P\bm{(}\log_{10}(\delta r);t\bm{)}$ for the A particles. The 
wider lines indicate increased time $t$.  a: $T=5.0$ (a) 
for $t = 0.01$, 0.05, 0.1, 5 and 10 listed from left to right. 
b: $T = 0.60$ for $t = 1.0$, 5, 15, 25 and 30 listed from
left to right. c: $T=0.47$ for
$t = 100$, 200, 300, 400, 500, and 800 listed from
left to right.  d: $T=0.44$ for $t = 1000$, 2000, 3000, 
4000, 5000, 6000, and 10000
listed from left to right.}
\end{figure*}
 
The values of the scaling exponents obtained from new fits 
for both the self-diffusion coefficient and the $\alpha$ relaxation time 
are quite a bit greater than those obtained from the conventional fit. Also, the
difference between the exponents for the A and B particles is considerably larger. 

We should point out that one could try using a slightly different procedure
to identify the mode-coupling transition temperature. Namely, one could try fit a straight
line to a plot of, \textit{e.g} $D^{1/\gamma}$ \textit{vs.} temperature, where $\gamma$
is a scaling exponent obtained from \emph{solving} mode-coupling equations. The are two
potentials problems with this approach: first, it requires solving full time and wave-vector
dependent mode-coupling equations \cite{FlennerSzamelMC}. Second, using this alternative approach 
one has to neglect the qualitative differences between the predictions of the theory
and results of the simulations like, \textit{e.g.} the difference between scaling exponents
for the A and the B particles and the difference between scaling exponents for the diffusion
coefficients and the relaxation times. 

In conclusion, we find that the mode-coupling
temperature of the Kob-Andersen binary mixture is not unique: if a temperature range different 
from the original one \cite{KobAndersen} is used for fitting 
the simulation data to the power laws, a different mode-coupling
transition temperature results. 
In the next section we show that there is a change in the dynamics 
around the commonly reported mode-coupling temperature, $T_c=0.435$.

\section{\label{distributions}Probability distributions of single particle 
displacements}

Following a procedure suggested previously
\cite{Puertas,Reichman,Cates}, we examined the
probability distributions of the logarithm of single-particle
displacements $P\bm{(}\log_{10}(\delta r);t\bm{)}$ at a time $t$.
Multiple peaks in $P\bm{(}\log_{10}(\delta r);t\bm{)}$
at a time $t^\prime$ have been clearly observed in simulations
of model colloidal gels \cite{Reichman}, and provide evidence of 
populations of fast and slow particles at $t^\prime$.   
Note that $P\bm{(}\log_{10}(\delta r);t\bm{)}$ is defined such
that the integral $\int_{x_0}^{x_1} P(x;t) \mbox{d}x$ is the
fraction of particles whose value of $\log_{10}(\delta r)$ is between
$x_0$ and $x_1$. Furthermore, the probability distribution 
$P\bm{(}\log_{10}(\delta r);t\bm{)}$ can be obtained from the self 
van Hove correlation function \cite{Hansen} since 
$P\bm{(}\log_{10}(\delta r);t\bm{)} = \ln(10) 4 \pi \delta r^3 G_s(\delta r,t)$.
There are some general properties of $P\bm{(}\log_{10}(\delta r);t\bm{)}$
if the van Hove correlation function is Gaussian. 
If the motion of a tagged particle is diffusive at all
times with a diffusion coefficient $D$, then the self
van Hove correlation function 
$G_s(\delta r,t) = (1/(4 \pi D t)^{\frac{3}{2}}) \exp(-\delta r^2/4 D t)$.
For a Gaussian van Hove function, the shape of $P\bm{(}\log_{10}(\delta r);t\bm{)}$
is independent of time \cite{note}, and the peak height of
$P\bm{(}\log_{10}(\delta r);t\bm{)} = \log_{e}(10) \sqrt{54/ \pi}\;e^{3/2} \approx 2.13$.
Deviations from this height represents deviations from Gaussian behavior of 
$G_s(r,t)$. 

\begin{figure*}
	\includegraphics[scale=0.25]{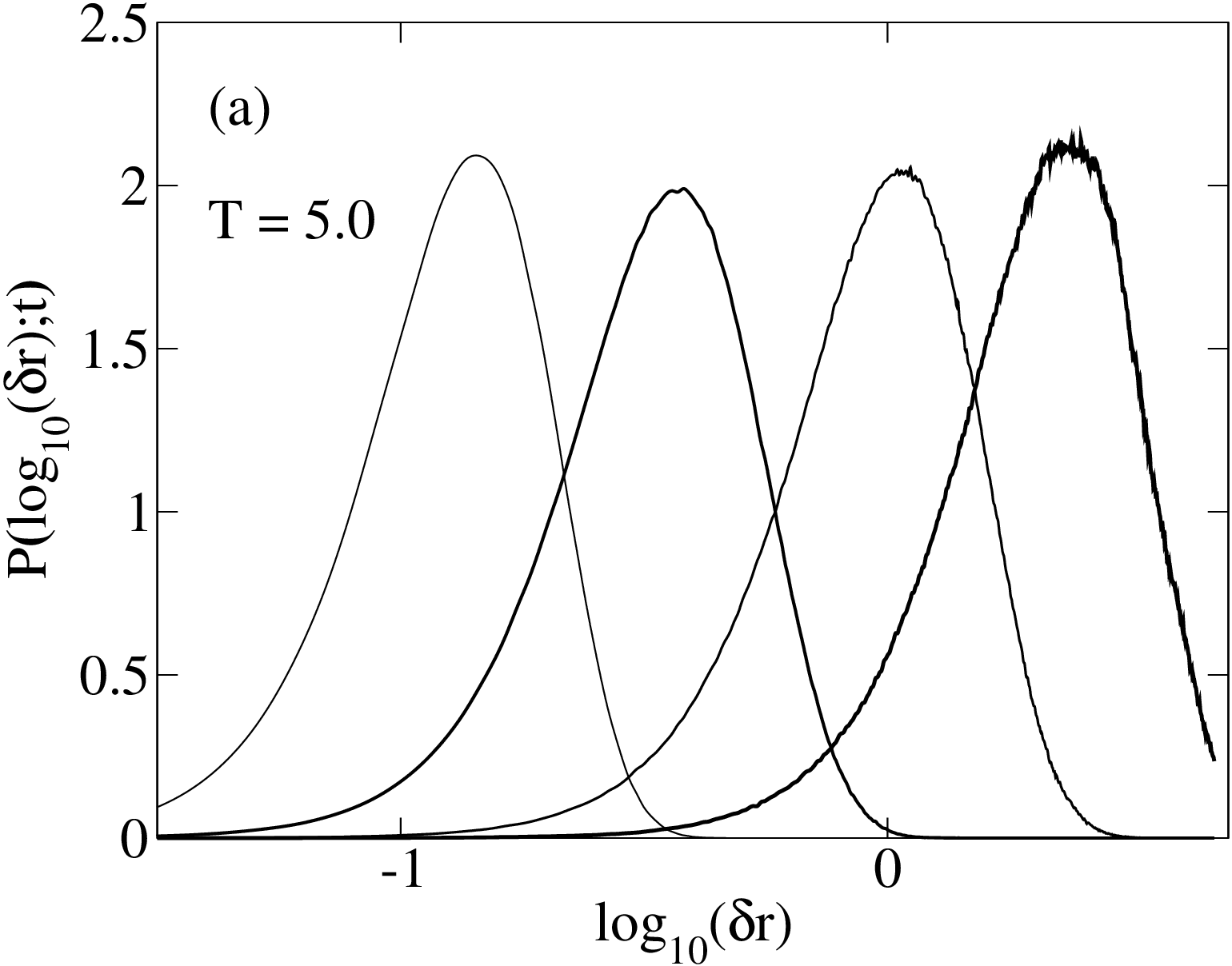}\hspace{1cm}
	\includegraphics[scale=0.25]{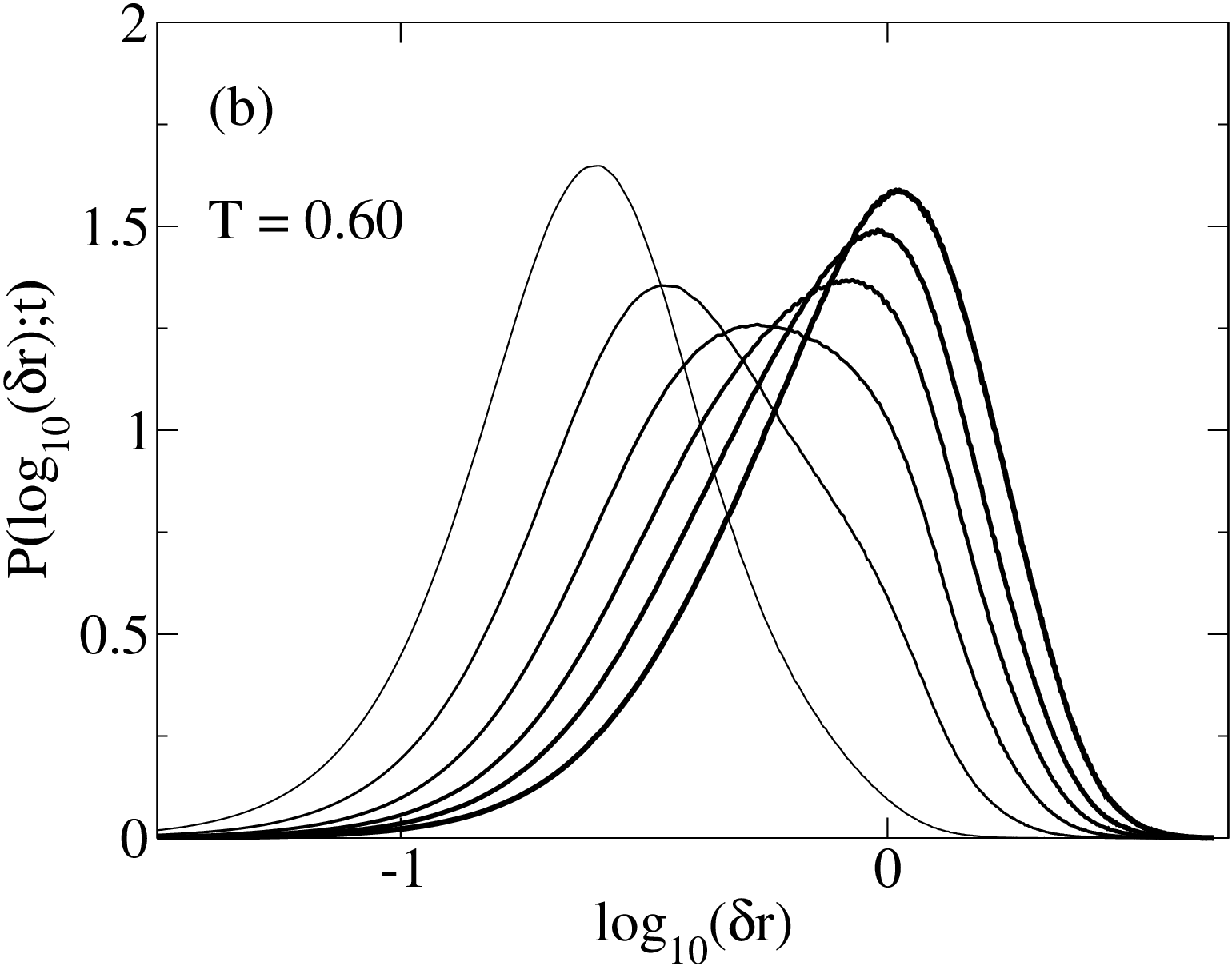}\\[0.25cm]
	\includegraphics[scale=0.25]{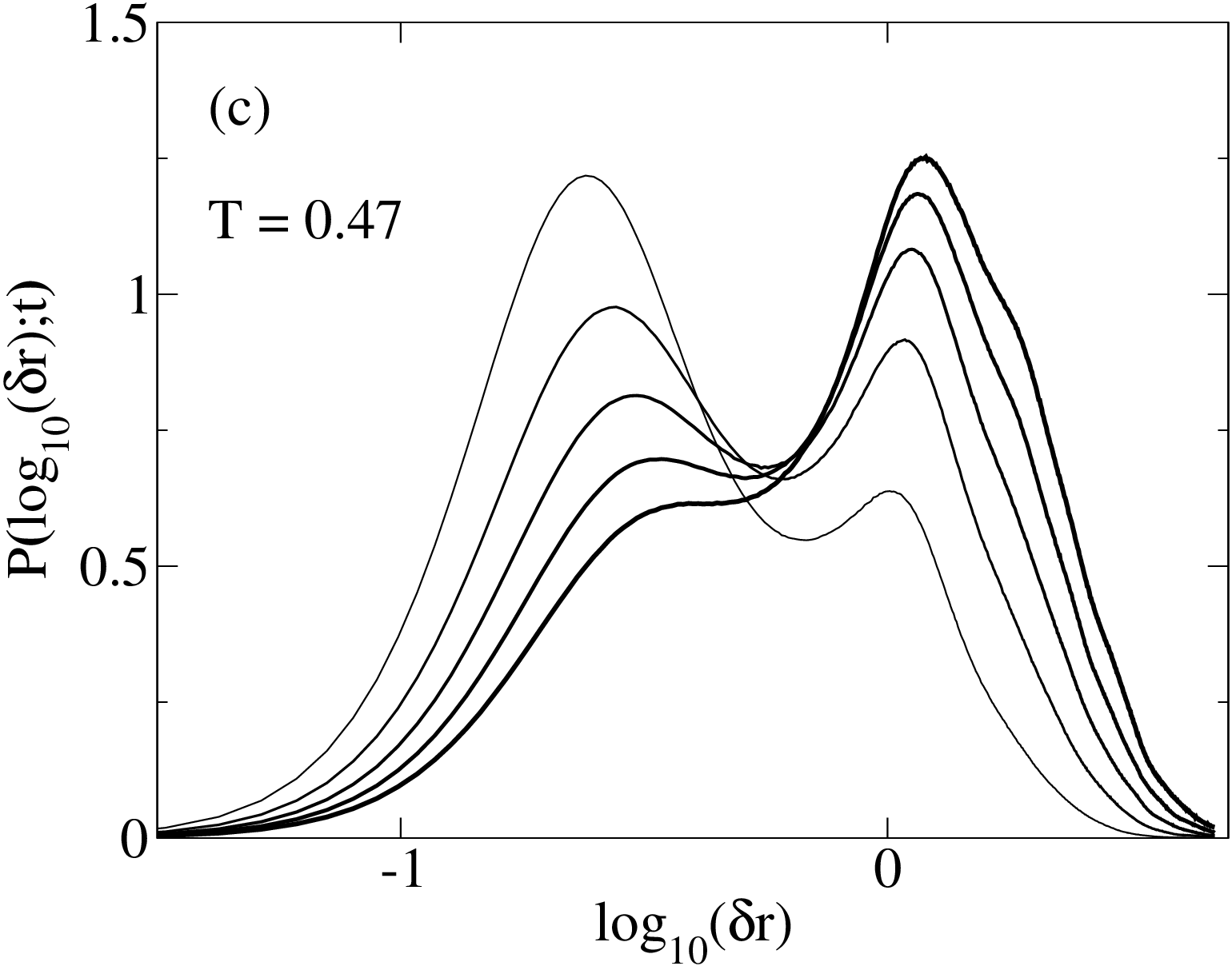}\hspace{1cm}
	\includegraphics[scale=0.25]{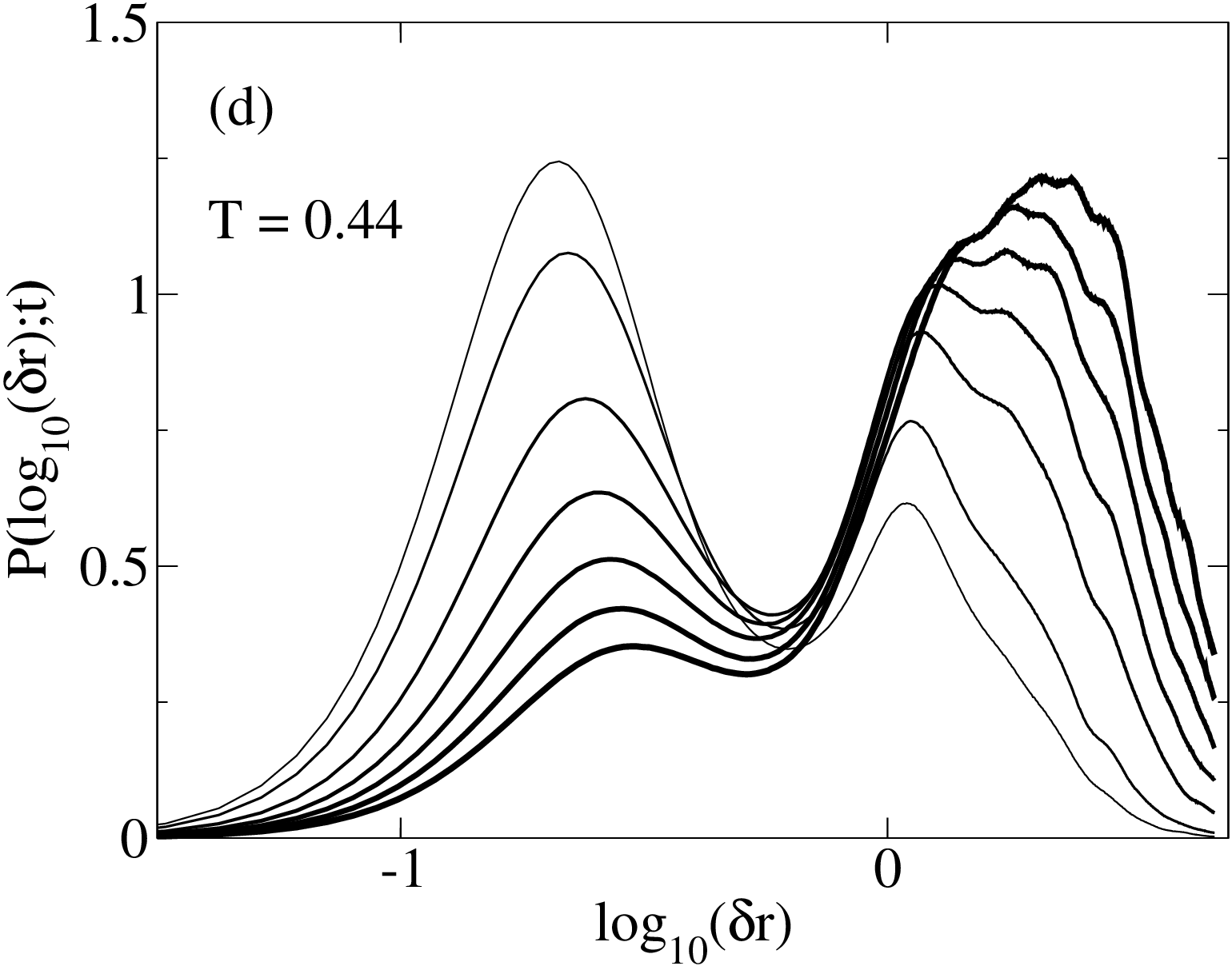}
\caption{\label{vhb}The probability of the logarithm of 
single particle displacements $P\bm{(}\log_{10}(\delta r);t\bm{)}$ for the B particles.
Wider lines indicates increased time $t$. (a): $T=0.5.0$ 
for $t = 0.01$, 0.05, 0.1, and 0.5 listed from left to right. 
(b): $T = 0.60$ for $t = 1$, 5, 10, 15, 20 and 25
listed from left to right. (c): $T = 0.47$ for
$t = 100$, 200, 300, 400, and 500 listed from left to right.  
(d): $T = 0.44$ for $t = 600$, 700, 800, 1000, 3000, 4000, 
and 5000 listed from left to right.}
\end{figure*}

Shown in Fig.~\ref{vha} is $P\bm{(}\log_{10}(\delta r);t\bm{)}$ for the 
A particles at $T=5.0$, 0.6, 0.47, and 0.44.  
For each temperature several different times are shown.  The 
thicker lines correspond to later times.  At $T=5.0$ there is little
deviation from Gaussian behavior: there is only one peak whose
height is close to 2.13 at all times. The peak position moves to
larger distances for larger times.  For $T=0.6$, there 
are deviations from Gaussian behavior manifested in the reduced
height of the peak of the distribution, but there is still
only one peak for all times.  At $T=0.47$, the deviations from Gaussian 
behavior are stronger and the distribution becomes very
broad at a time which corresponds to right after the 
plateau of the mean square displacement.  
At $T=0.44$ there are two distinct peaks.
The position of the second peak depends on time, but when 
the height of both peaks are approximately equal the position of the second peak is 
around $\log_{10}(\delta r) = 0$, thus $\delta r = 1.0 = \sigma_{AA}$. 

The probability distributions $P\bm{(}\log_{10}(\delta r);t\bm{)}$ provide clear evidence
that there are populations of particles
with different mobilities.  These probability distributions 
are similar to the distributions observed in model colloidal gels close to a
gelation transition \cite{Puertas} and in dense systems with purely repulsive 
interactions \cite{Reichman}. The shape of the distributions strongly suggests
a heterogeneous hopping-like motion for at least a fraction of the
particles.  We find that the hopping rate of the particles varies 
greatly between particles, and the typical size
of the hopping length for the A (larger) particles 
is equal to one particle diameter.

The two peaks in $P\bm{(}\log_{10}(\delta r);t\bm{)}$ 
are clearly defined at a higher temperature
for the B particles, Fig.~\ref{vhb}, than for the A particles.
For $T=5.0$, $P\bm{(}\log_{10}(\delta r);t\bm{)}$ for the B particles 
is similar to what is observed
for the A particles.  The distribution $P\bm{(}\log_{10}(\delta r);t\bm{)}$
broadens and the peak height decreases significantly
for the B particles starting at $T=0.6$, and we observe the two peak
structure starting at $T=0.5$. The two peaks are well defined for
$T=0.47$ and are very prominent for $T=0.44$. At these low temperatures there is a clear
distinction between mobile and immobile particles. The second peak 
occurs around $\log_{10}(\delta r) \approx 0.086$ 
for the B particles, which
corresponds to a value of $\delta r \approx 1.25$.
This suggests that the typical length of the particle
jumps are slightly larger
for the B particles than for the A particles.
Note that there are small
secondary peaks in $P\bm{(}\log_{10}(\delta r);t\bm{)}$ for the B
particles for the longer times.  This is expected
if the activated hopping is the dominant relaxation process.  

The size dependence of the distributions of the displacements of the particles 
have been observed in simulations of colloidal gels by 
Puertas \textit{et al.}~\cite{Puertas}.
They simulated a system of soft core polydisperse particles with an average
particle radius $a$.  They noticed that the distributions of squared
displacements were more bimodal for particles with a smaller radius 
(see Fig.~6 in Ref.~\cite{Puertas}). It is likely that the size dependence
of dynamic heterogeneity is  a general feature of systems with slow dynamics.

In earlier studies \cite{Puertas,Reichman} probability distributions
of particle displacements were examined at 
a time $t^*$ such that $\left< \delta r^2(t^*) \right> = 10 a^2$, where $a$
is a measure of the average size of the polydisperse particles. 
While bimodal distributions were observed for this $t^*$,
there was no clear justification for selecting this particular timescale.
In the next section we 
introduce a new non-Gaussian parameter whose peak position
allows us to identify the time in which the two peaks in 
$P\bm{(}\log_{10}(\delta r);t\bm{)}$ are around the same height.  
Moreover, we will demonstrate that the peak position of the commonly
used non-Gaussian parameter $\alpha_2(t)$ does a poor job of identifying
this time.  

\section{\label{newnong}New non-Gaussian parameter}

Many simulations which examine heterogeneous dynamics 
focus on a timescale which is found from the peak of the
non-Gaussian parameter 
$\alpha_2(t) = \frac{3}{5} \left< \delta r^4(t) \right>/\left< \delta r^2(t) \right>^2 - 1$.  
It has been observed that close to $T_c = 0.435$, 
$\alpha_2(t)$ has a peak occurring at the so-called late $\beta$ regime, 
\textit{i.e.}~some
time before the $\alpha$ relaxation time. Moreover, for $T< 0.8$ with decreasing temperature 
the value of the non-Gaussian parameter at the $\alpha$ relaxation time, $\alpha_2(\tau_{\alpha})$, 
is a decreasing 
fraction of its maximum value \cite{SzamelFlenner}.
In particular, at $T=0.44$   the non-Gaussian parameter at the $\alpha$ relaxation time 
is approximately equal to 0.62 and 0.35 of its
maximum value for the A and B particles, respectively.
We observe that the two peaks
in $P\bm{(}\log_{10}(\delta r);t\bm{)}$ (for temperatures in which the two peaks
are clearly defined) are about the same height at some time
after the $\alpha$ relaxation time.  Thus $\alpha_2(t)$ is
small compared to its maximum value at a time when the two peaks are very prominent. 
 
\begin{figure}
	\includegraphics[scale=0.25]{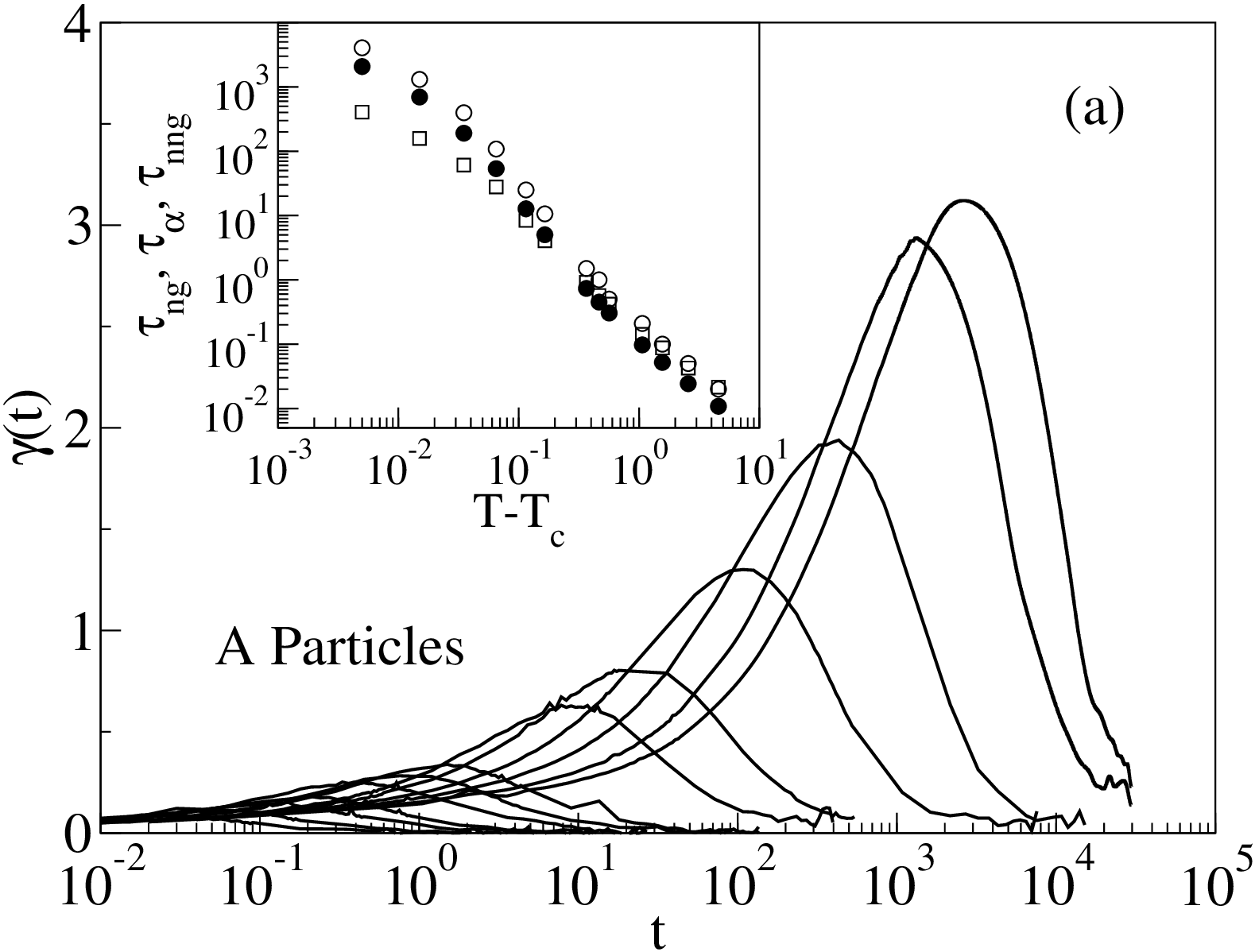}\\[0.25cm]
	\includegraphics[scale=0.25]{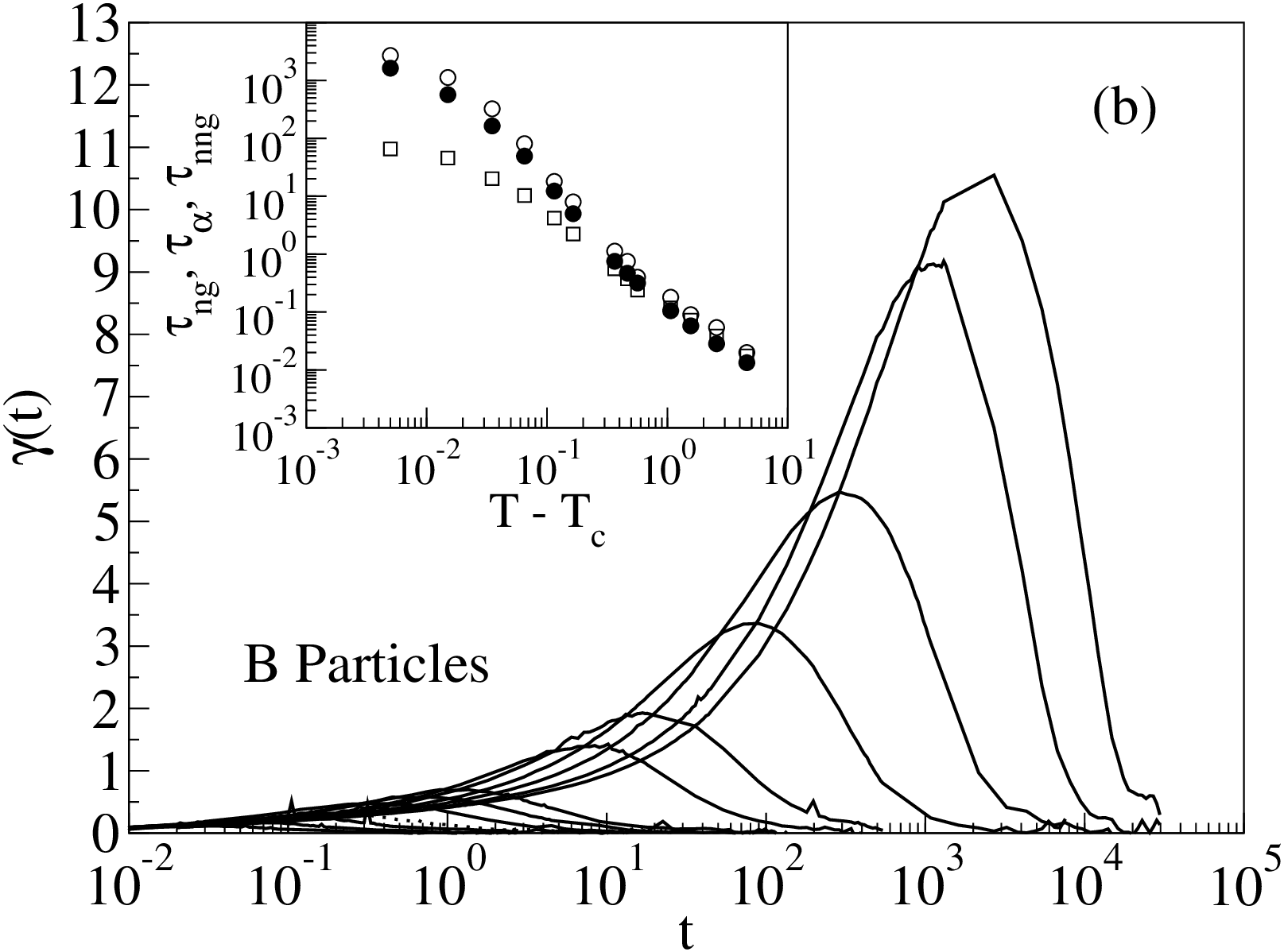}
\caption{\label{newng}The new non-Gaussian parameter $\gamma(t)$ for 
$T=5.0$, 3.0, 2.0, 1.5, 1.0, 0.9, 0.8, 0.6, 0.55, 0.50, 0.47, 0.45, 0.44 for
the A (a) and B (b) particles.  The insets are the
position of the peak of $\gamma(t)$, (open circles)
compared to $\alpha$ relaxation time (closed circles) and
the peak position of $\alpha_2(t)$ (squares) versus temperature.}
\end{figure}

We propose to use a new function that, as does $\alpha_2(t)$, quantifies deviations from
a Gaussian distribution of displacements, 
\begin{equation}
\gamma(t) = \frac{1}{3} \left\langle \delta r^2 \right\rangle 
\left\langle \frac{1}{\delta r^2} \right\rangle - 1
\label{gamma},
\end{equation}
where $\delta r$ is the distance over which the particle moved in time $t$.
We will show that this new non-Gaussian parameter 
identifies the time in which the two peaks in the probability distribution
of the logarithm of single-particle displacements are most evident.
The parameter $\alpha_2(t)$ is significantly influenced mostly by
particles which move farther than expected from a Gaussian
distribution of particle displacements.  
In contrast, the parameter $\gamma(t)$ weights very strongly the particles which
have not moved as far as expected from a Gaussian distribution of 
particles displacements through the term 
$\left \langle 1/\delta r^2 \right \rangle$ 
and weights the particles which move farther
than expected from a Gaussian distribution through the 
$\left\langle \delta r^2 \right \rangle$ term.  
The factor of $1/3$ ensures that $\gamma(t)$ is zero when 
the self part of the van Hove correlation function is Gaussian.    

Shown in Fig.~\ref{newng} is $\gamma(t)$ for the A and
B particles.  At short times, the motion of the
particles are Gaussian and $\gamma(t)$ is close to zero.
At intermediate times there is a peak in $\gamma(t)$ whose 
height and position increases with decreasing temperature.  
The peak in $\gamma(t)$ is larger for the B particles, which is 
what we expect from examining the self part of the van Hove correlation function
and $P\bm{(}\log_{10}(\delta r);t\bm{)}$. At long times
$\gamma(t)$ decays to zero.  

It is interesting to note the temperature dependence of the 
the average mean square displacement at the peak position of $\gamma(t)$:
for $T \ge 0.8$ the average mean square displacement is independent of temperature and 
it is around $0.23 \sigma_{AA}^2$ for the A particles 
and $0.31 \sigma_{AA}^2$ for the B
particles.  For $T < 0.8$, the average mean squared displacement at the peak
position of $\gamma(t)$ increases with decreasing temperature and at
$T = 0.44$ it reaches the values of approximately
$0.84 \sigma_{AA}^2$ for the A particles and 
$2.40 \sigma_{AA}^2$ for the B particles.  It should be pointed out that the
temperature at which the average mean squared displacement at the peak
position of $\gamma(t)$ starts increasing is close to the so-called onset
temperature identified for the Kob-Andersen model by Brumer and Reichman \cite{onset}.

\begin{figure}
	\includegraphics[scale=0.25]{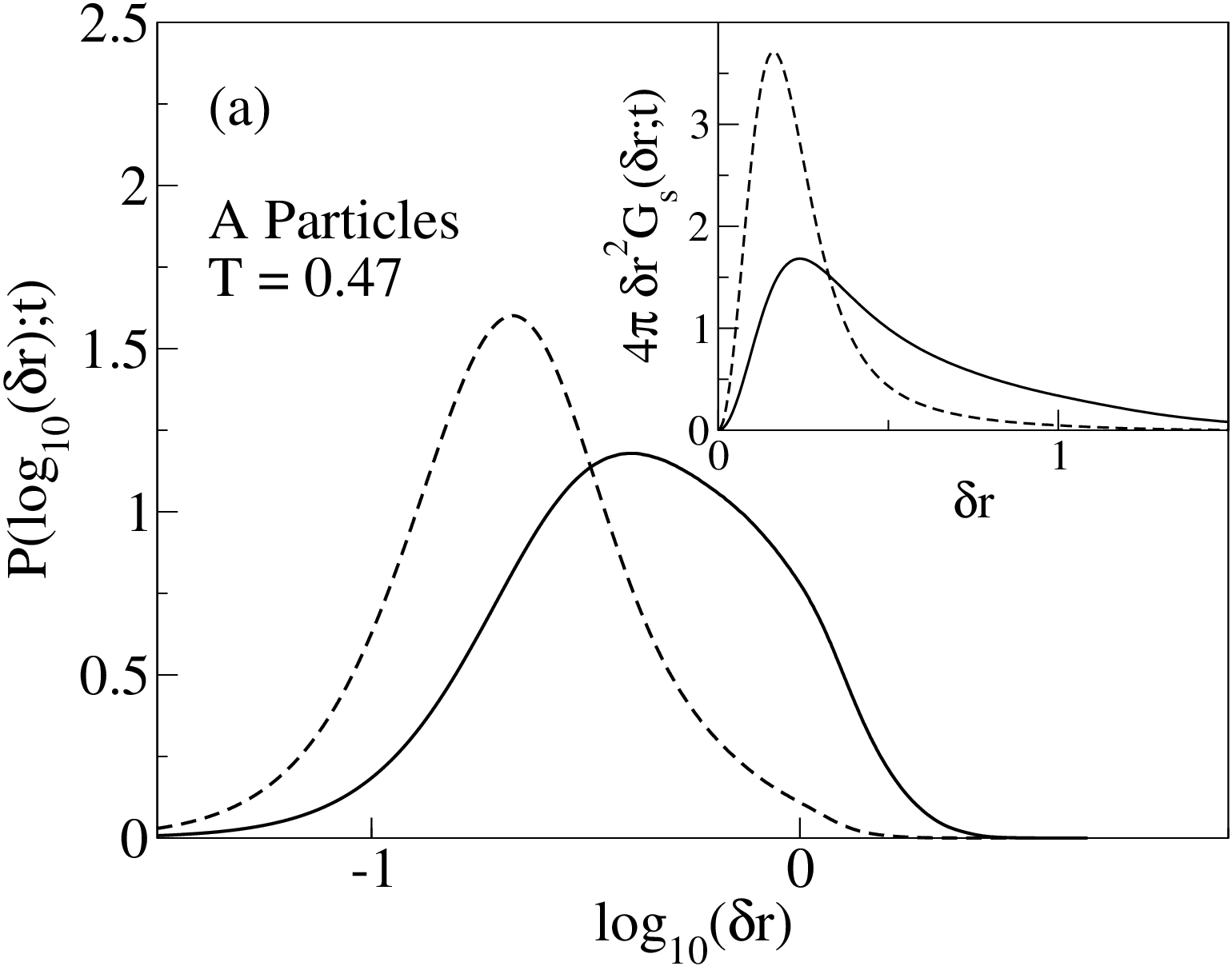}\\[0.25cm]
	\includegraphics[scale=0.25]{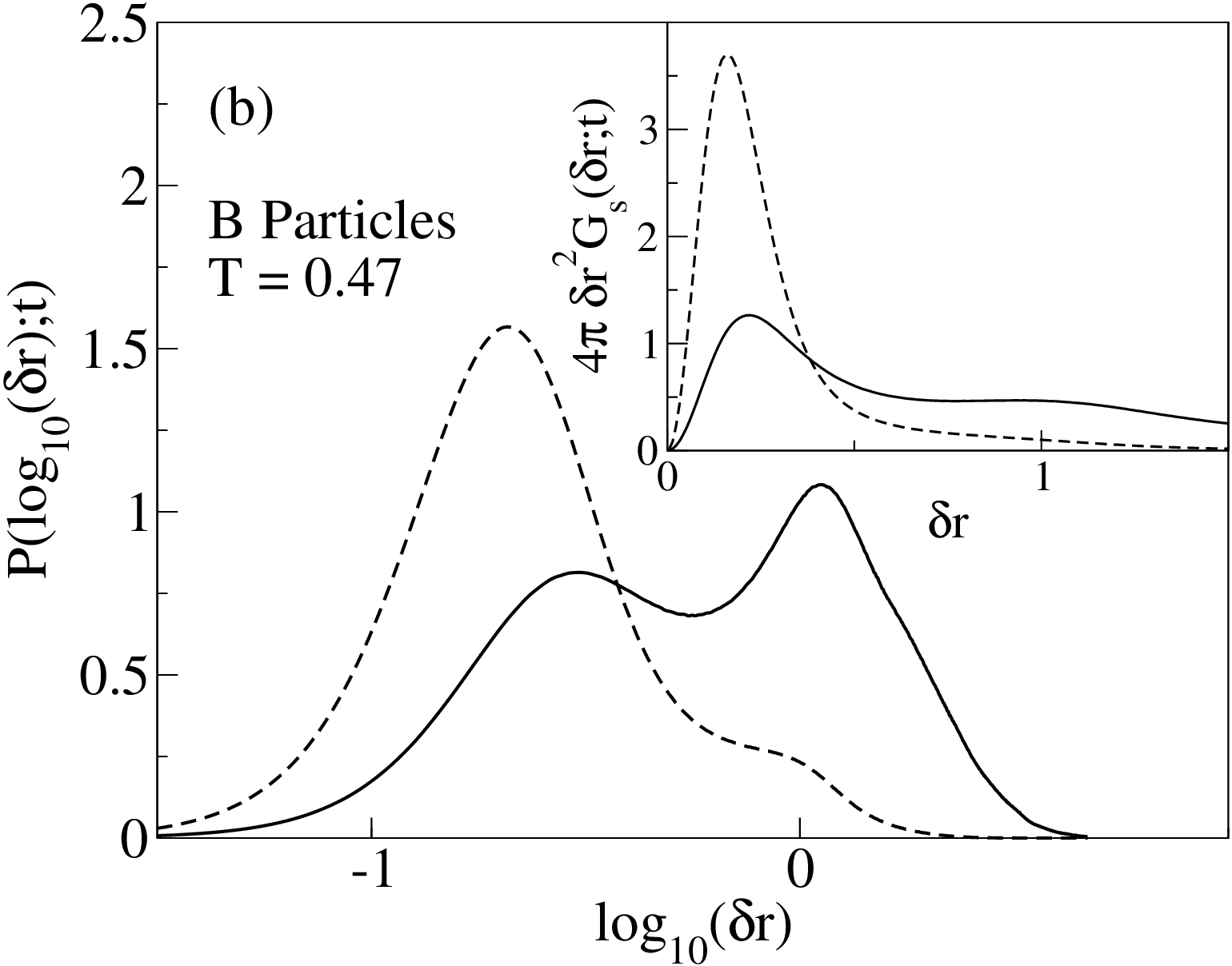}
\caption{\label{comp}The probability of the logarithm of the displacements
at the the peak position of the the old non-Gaussian parameter $\alpha_2(t)$
(dashed line) and the new non-Gaussian parameter $\gamma(t)$ 
(solid line) at $T = 0.47$ 
for the A particles (a) and the B particles (b).  The inserts 
are the self part of the van Hove correlation functions at the peak
position of $\alpha_2(t)$ (dashed line) and $\gamma(t)$ (solid line).}
\end{figure}

The insert to each graph in Fig.~\ref{newng} is the peak position of the new non-Gaussian parameter,
$\gamma(t)$, denoted by $\tau_{nng}$ compared to the $\alpha$ relaxation time and
the peak position of the commonly used non-Gaussian parameter, $\alpha_2(t)$, denoted by $\tau_{ng}$.  
Notice that the peak position of $\gamma(t)$ is always greater than the $\alpha$ relaxation time,
but it has the same temperature dependence.  This is in
contrast to $\tau_{ng}$ which is
greater than the $\alpha$ relaxation time and 
equal to $\tau_{nng}$ at higher temperatures, but increases slower
with decreasing temperature than $\tau_{\alpha}$ and
$\tau_{nng}$.  For the lowest temperatures, $\tau_{ng}$ is
much smaller than either $\tau_{\alpha}$ and $\tau_{nng}$.

\begin{figure}
	\includegraphics[scale=0.25]{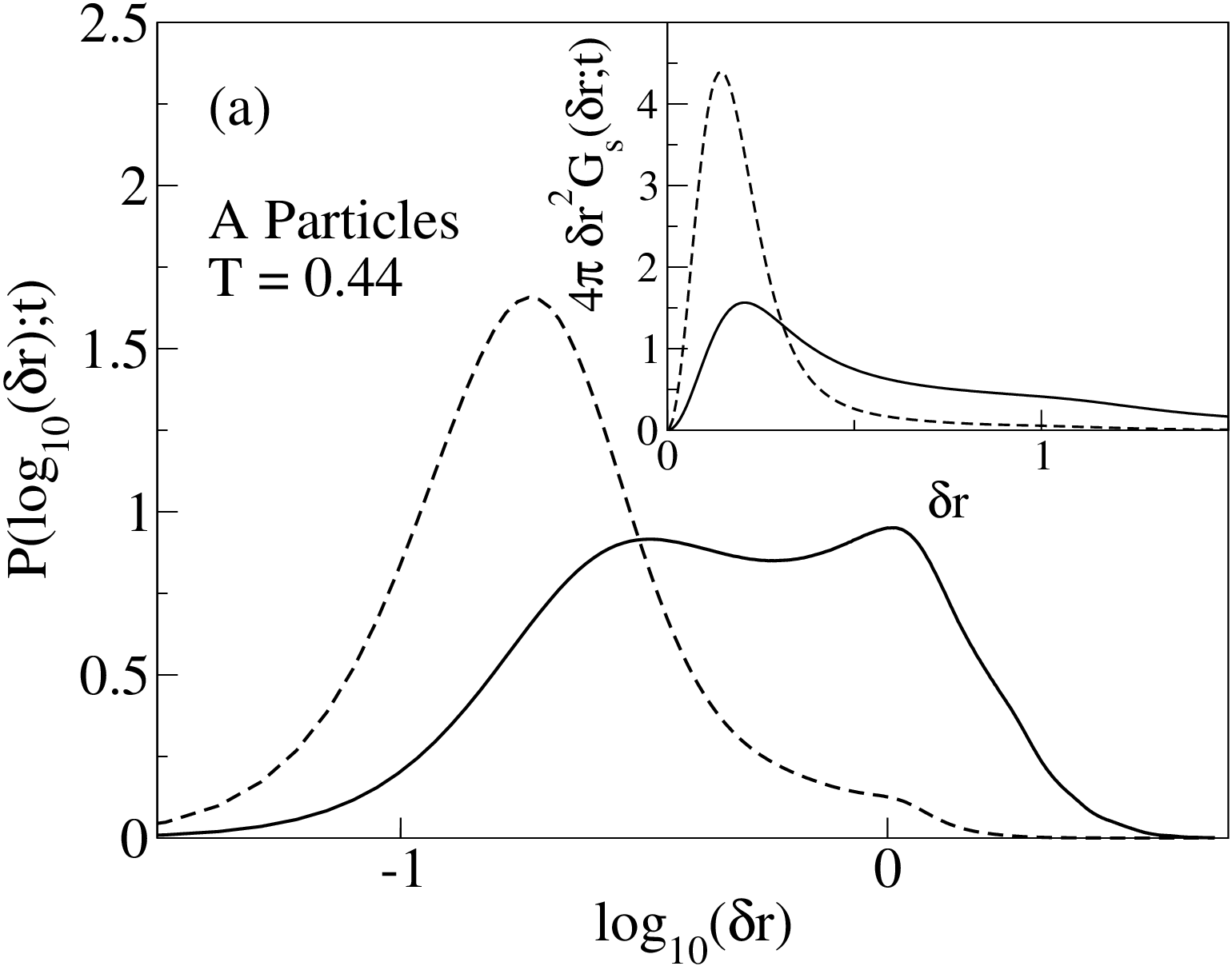}\\[0.25cm]
	\includegraphics[scale=0.25]{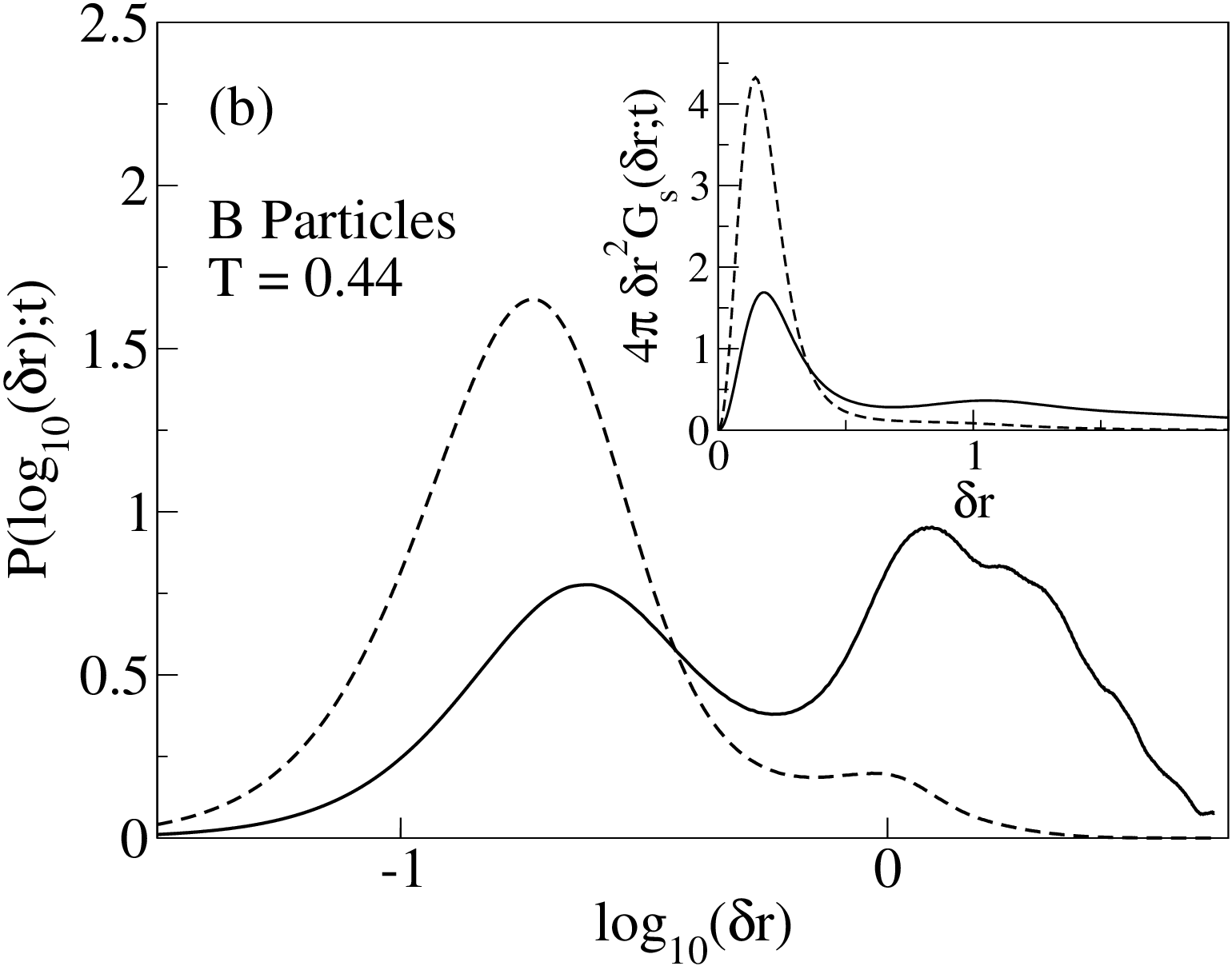}
\caption{\label{comp1}The probability of the logarithm of the displacements
at the peak position of the old non-Gaussian parameter $\alpha_2(t)$ (dashed line)
and the new non-Gaussian parameter $\gamma(t)$ (solid line) 
for $T=0.44$ for the A particles (a)
and the B particles (b).  The inserts are the self part of the van Hove correlation functions at
the peak position of $\alpha_2(t)$ (dashed line) and $\gamma(t)$ (solid line).}
\end{figure}

The peak position approximately corresponds to the
time in which the two peaks of $P\bm{(}\log_{10}(\delta r);t\bm{)}$ are of equal height.
Shown in Fig.~\ref{comp} and Fig.~\ref{comp1} is
the $P\bm{(}\log_{10}(\delta r);t\bm{)}$ for the A and B particles at
the peak position of $\alpha_2(t)$ (dashed lines) and
$\gamma(t)$ (solid lines) for temperatures of $T = 0.47$ and
$T = 0.44$.  Notice that at the time corresponding to the peak 
position of $\alpha_2(t)$, there is at most a shoulder in
$P\bm{(}\log_{10}(\delta r);t\bm{)}$.  For the time which corresponds
to the position of the peak in $\gamma(t)$, the two 
peaks in $P\bm{(}\log{10}(\delta r);t\bm{)}$ are of similar height and
the definition of mobile and immobile particles is clear.
For comparison,
the self part of the van Hove correlation function at the peak 
position of $\alpha_2(t)$ (dashed lines) and $\gamma(t)$
(solid lines) are shown as inserts in Fig.~\ref{comp} and 
Fig.~\ref{comp1}. 

\section{\label{conclusion}Conclusions}

The mode-coupling theory has been used extensively to describe
the relaxation in supercooled liquids and the glass transition.
It correctly describes many qualitative features of the glass
transitions observed experimentally and in computer simulations.
The most notable success of the mode-coupling theory is that
it correctly describes the two step decay of the intermediate
scattering functions and the qualitative features of the
mean square displacement as a function of time.  

The idealized mode-coupling theory predicts power law divergence
of the $\alpha$ relaxation time and power law
vanishing of the self-diffusion coefficient
at a transition temperature $T_c$.  It has been observed that
there is power law like behavior of the $\alpha$ relaxation
time and the diffusion coefficient in simulations and experiments
close to a temperature $T_c$. However, the structural arrest 
predicted by the mode-coupling theory does not occur at $T_c$, but
rather there appears to be a crossover to a different relaxation scenario
and the mode-coupling transition is said to be ``avoided''.  

We performed Brownian dynamics simulations of a frequently studied
glass forming binary mixture for a large range of temperatures.
The temperature dependence of the $\alpha$ relaxation time
and the diffusion coefficient were similar to what was 
observed in previous simulations of the same system 
\cite{KobAndersen,Kobsr,Gleim}.
We found that 
the diffusion coefficient and the $\alpha$ relaxation time
can also be fit to a power law where the transition temperature is 
$T_c^{(2)} = 0.401$ which is lower than the generally accepted 
mode-coupling transition temperature $T_c = 0.435$.  However, by
examining the probability distributions of the logarithm of single
particle displacements, $P\bm{(}\log_{10}(\delta r);t\bm{)}$, we demonstrated
that the crossover from the high temperature diffusive relaxation
of the particles to low temperature hopping-like motion occurs near
the generally accepted 
mode-coupling transition temperature $T_c = 0.435$.

The change in the relaxation processes are evident when one
examines the distribution of the logarithm of single
particle displacements $P\bm{(}\log_{10}(\delta r);t\bm{)}$. 
At higher temperatures there is one peak in $P\bm{(}\log_{10}(\delta r);t\bm{)}$
whose position increases for increased time.    
For lower temperatures, 
the distribution becomes broad at a time scale right after the 
plateau region of the mean square displacement.  At the
lowest temperatures examined in this study, there
are two distinct peaks in $P\bm{(}\log_{10}(\delta r);t\bm{)}$ for
both the A and B particles.  The two peaks are 
evidence that on an intermediate time scale the particles can be separated by their individual
relaxation time, and thus are dynamically heterogeneous.  Moreover, the
minimum between the peaks is smaller for the B
particles than for the A particles at a fixed temperature.  The 
dependence on particles size of the probability distributions is similar
to what was observed in earlier simulations \cite{Puertas,Reichman}.

One possible interpretation of our results could be that the mode-coupling temperature
does not have any physical significance and the crossover in supercooled liquid's
dynamics is very smooth \cite{Berthier}. While such an interpretation cannot be excluded, 
we would like to advocate a more cautious conclusion: in order to 
identify a crossover temperature (or a narrow crossover temperature \emph{range}) that one could
interpret as mode-coupling temperature, one has to investigate
not only power-law fits to transport coefficients and/or relaxation times but also whether
microscopic dynamics is homogeneous and diffusive-like or heterogeneous and hopping-like. 
We would like to emphasize that although the mode-coupling theory, in its standard form,
cannot describe hopping-like dynamics, it does provide a reasonable \cite{FlennerSzamelMC}
description of dynamics in moderately supercooled fluids.

We found that the typical non-Gaussian parameter $\alpha_2(t)$
does a poor job of identifying the time scale on which the heterogeneous, hopping-like 
motion is most evident.  We defined a new non-Gaussian parameter 
$\gamma(t) = \frac{1}{3}\left< \delta r^2 \right> \left< 1/\delta r^2 \right> - 1$.
For temperatures in which there are two peaks in $P\bm{(}\log_{10}(\delta r);t\bm{)}$,
$\gamma(t)$ has a peak occurring at a time
$\tau_{nng}$ in which the two peaks have approximately
the same height.  For times in which only one
peak is present, the peak position of $\gamma(t)$ identifies the time in 
which $P\bm{(}\log_{10}(\delta r);t\bm{)}$ is widest.  The position of the peak 
for $\alpha_2(t)$ and $\gamma(t)$ are the same for high temperatures,
but the peak position of $\alpha_2(t)$ increases slower with decreasing
temperature than the peak position of $\gamma(t)$.  Moreover, the peak 
position of $\gamma(t)$ has the same temperature dependence as the
$\alpha$ relaxation time.

\section*{Acknowledgments}
We gratefully acknowledge the support of NSF Grant No.~CHE 0111152.


\begin{thebibliography}{99}
\bibitem{note} If the van Hove correlation function is Gaussian, all the central moments 
of the probability distribution of the logarithm of single-particle displacements,
$\int_{-\infty}^{\infty}\left( x - \bar{x} \right)^n P(x;t)\, \mbox{d}x$, where
$\bar{x} =  \int_{-\infty}^{\infty} x P(x;t)\, \mbox{d}x$,
are independent of time.
\end{thebibliography}
\end{document}